\documentclass[11pt, a4paper]{article}
\usepackage{amsmath,amssymb}
\newtheorem{theorem}{Theorem}

\begin{document}

\title{Second order quasilinear PDEs  and  conformal structures in projective space}
\author{P.A. Burovskiy, E.V. Ferapontov \  and
S.P. Tsarev\thanks{SPT acknowledges partial financial support from
the RFBR grant 06-01-00814-a and the Russian--Taiwanese grant
06-01-89507-HHC (95WFE0300007).} }
   \date{}
\vspace{-20mm}
   \maketitle
\vspace{-7mm}
\begin{center}
Department of Mathematical Sciences \\ Loughborough
University \\
Loughborough, Leicestershire LE11 3TU, UK \\[1ex] 
and \\[1ex]
Institute of Mathematics\\
Siberian Federal University\\
79 Svobodny Prospect\\
Krasnoyarsk 660041 Russia\\[1ex]
e-mails: \\
\texttt{P.A.Burovskiy@lboro.ac.uk}\\
\texttt{E.V.Ferapontov@lboro.ac.uk}\\
\texttt{tsarev@newmail.ru}\\
\end{center}

\medskip

\begin{abstract}
We investigate second order quasilinear equations of the form
$$
f_{ij} u_{x_ix_j}=0,
$$
where $u$ is a function of $n$ independent variables $x_1, ..., x_n$, and the coefficients $f_{ij}$ depend on the first order derivatives $p^1=u_{x_1}, ...,  p^n=u_{x_n}$ only. We demonstrate that the natural equivalence group of the problem is isomorphic to $SL(n+1, R)$, which  acts by projective transformations on the space $P^n$ with coordinates $p^1, ...,  p^n$. The coefficient matrix $f_{ij}$ defines on $P^n$ a conformal structure $f_{ij}({\bf  p}) dp^idp^j$.
In this paper we concentrate on the case $n=3$, although some  results hold in any dimension.
The necessary
and sufficient conditions for the integrability of such equations by  the method of
hydrodynamic reductions are derived. These conditions constitute an over-determined system of PDEs for the coefficients  $f_{ij}$,  which is in involution. We prove that the moduli space of integrable equations is $20$-dimensional. Based on these results, we show that any equation satisfying the integrability conditions is necessarily conservative, and possesses a dispersionless Lax pair. Reformulated in differential-geometric terms,  the integrability conditions imply that the  conformal structure $f_{ij}({\bf  p}) dp^idp^j$ is conformally flat, and possesses an infinity of $3$-conjugate null coordinate systems parametrized by three arbitrary functions of one variable. Integrable equations provide an abundance of explicit examples of such conformal structures parametrized by elementary functions, elliptic functions and modular forms.

\medskip


\medskip

\end{abstract}

\newpage

 \section{Introduction}

The main object of our study are  second order  quasilinear  equations of the form
 \begin{equation}
f_{11} u_{xx} + f_{22} u_{yy} + f_{33} u_{tt} + 2 f_{12} u_{xy} + 2f_{13} u_{xt} + 2f_{23} u_{yt} =0,
 \label{u}
 \end{equation}
where $u$ is a function of three independent variables $x, y, t$, and the coefficients
$f_{ij}$ depend on  the first order derivatives  $u_x, u_y, u_t$ only. Equations of this type
arise in a wide range of applications in mechanics, general relativity, differential geometry
and the theory of integrable systems.
Among the most familiar examples  one should  mention the Boyer-Finley equation,
$$
u_{xx}+u_{yy}-e^{u_t}u_{tt}=0,
$$
which is descriptive of a class of self-dual $4$-manifolds \cite{BF}, as well as the dispersionless Kadomtsev-Petviashvili (dKP) equation,
$$
u_{xt}-u_xu_{xx}-u_{yy}=0,
$$
also known as the Khokhlov-Zabolotskaya equation, which arises in  non-linear acoustics \cite{KZ} and  the theory of Einstein-Weyl structures \cite{D}.

The integrability aspects of equations of the  form (\ref{u}) have been investigated by a whole variety of modern techniques including symmetry analysis, differential-geometric and algebro-geometric methods, dispersionless $\bar \partial$-dressing, factorization techniques, Virasoro constraints, hydrodynamic reductions, etc. However, until recently there was no intrinsic  approach which would allow a unified treatment of these and other examples.  Moreover, there was no satisfactory definition of the integrability  which would (a) be algorithmically verifiable,  (b) lead to  classification results and (c) provide a scheme for the construction of exact solutions. We emphasize that equations of the form (\ref{u}) differ essentially from their solitonic counterparts, and require an alternative approach.
Such approach, based on the method of hydrodynamic reductions and, primarily,  on the work \cite{GibTsa96, GibTsa99}, see also \cite{Kodama, Kr1, Kr, F, Fer22, Ma}, etc, was proposed in \cite{Fer4, FKT}. It was suggested to define the integrability of a multi-dimensional dispersionless system by requiring the existence of `sufficiently many' hydrodynamic reductions which provide multi-phase solutions playing a role similar to that of algebro-geometric solutions of soliton equations.

In Sect. 2 we briefly review the method of hydrodynamic reductions, and apply it to equations of the form (\ref{u}). This leads to a system of differential constraints for the coefficients $f_{ij}$, which are necessary and sufficient for the integrability.
We demonstrate that the system of constraints is in involution, and prove our first main result (Theorem 1 of Sect. 2):

\begin{itemize}
\item {\bf The moduli space of integrable equations of the form (\ref{u}) is $20$-dimensional. }
\end{itemize}

 In Sect. 3 we point out that the class  of equations (\ref{u}) is form-invariant under the action of $SL(4, R)$, which  constitutes the equivalence group of the problem. This action  corresponds to linear transformations of the variables $x, y, t$ and $ u$. Since equivalence transformations preserve the integrability, any two $SL(4, R)$-related equations are regarded as `the same'. All our classification results   are obtained modulo this equivalence.

 Based on the integrability conditions, in Sect. 4 we  classify integrable equations of the form (\ref{u}) under various simplifying assumptions. This leads to a wide class of non-trivial examples, both known and new, which are expressible in terms of elementary functions, elliptic functions and modular forms. Just to mention a few of them, we have found an integrable equation
$$
\alpha \frac{\wp'(u_x)-\wp'(u_y)}{\wp(u_x)\wp(u_y)}\ u_{xy}+
\beta \frac{\wp'(u_t)-\wp'(u_x)}{\wp(u_x)\wp(u_t)}\ u_{xt}+
\gamma \frac{\wp'(u_y)-\wp'(u_t)}{\wp(u_y)\wp(u_t)}\ u_{yt}=0,
$$
here $\wp$ is the Weierstrass $\wp$-function, $(\wp')^2=4\wp^3-g_3$, and $\alpha, \beta, \gamma$ are arbitrary constants.
Another interesting example comes from the class
$$
u_{xy}+(u_xu_yr(u_t))_t=0;
$$
for equations of this form the integrability conditions result in a single third order ODE for $r$,
\begin{equation*}
r'''(r'-r^2) -r''^2+ 4{r}^3{r''}+2r'^3 - 6{r}^2{r'}^2= 0,
\end{equation*}
which  appeared recently in the context   of  modular forms of level two \cite{A2}. Its generic solution is given by the  Eisenstein series
$$
{\cal E}(q)=1-8\sum_{n=1}^{\infty}\frac{(-1)^nnq^n}{1-q^n}, ~~~ q= e^{4u_t},
$$
which is associated with the congruence subgroup $\Gamma_0(2)$ of the modular group. Further examples of integrable equations expressible in terms of modular forms can be found in  \cite{Odesskii} in the classification of integrable Lagrangian equations of the form (\ref{u}) corresponding to  first order Lagrangian densities $g(u_u, u_y, u_t)$.  A generic  Lagrangian density turns out to be a modular form of its arguments.
Further generalizations of the above example include the equation
\begin{equation*}
 u_{xy} + 2\left( u_x\, (\log \theta)' \right)_t = 0,
\end{equation*}
where $\theta\left(\frac{u_y}{2\pi},-\frac{u_t}{\pi i} \right)$ is the Jacobi theta-function,
\begin{equation*}
  \theta(z,\tau) = 1 + 2\sum_{n=1}^\infty e^{\pi i n^2 \tau} \cos(2\pi n z),
\end{equation*}
and prime denotes differentiation by $u_y$.

In Sect. 5 we study first order conservation laws, that is, relations of the form
\begin{equation}\label{conserv-law}
    g_1(u_x,u_y,u_t)_x + g_2(u_x,u_y,u_t)_y + g_3(u_x,u_y,u_t)_t = 0,
\end{equation}
which hold identically modulo (\ref{u}). Our second result (Theorem 2 of sect. 5) states that

\begin{itemize}
\item {\bf Any integrable equation of the form (\ref{u})  possesses exactly four first order conservation laws. }
\end{itemize}

In Sect. 6 we investigate the existence of  dispersionless Lax pairs (scalar pseudo-potentials),
\begin{equation}
S_t=F(S_x, \, u_x, \, u_y, \, u_t), ~~~
S_y=G(S_x, \, u_x, \, u_y, \, u_t),
\label{Lax}
\end{equation}
which imply Eq. (\ref{u}) via the consistency condition $S_{ty}=S_{yt}$ (we point out that the dependence of $F$ and $G$ on $S_x$ is generally non-linear). Lax pairs of this type first appeared in the construction of the universal Whitham hierarchy, see \cite{Kr, Kr3} and references therein.
 It was observed in \cite{Zakharov} that  consistent Hamilton-Jacobi-type relations of the form (\ref{Lax}) arise from the usual `solitonic'  Lax pairs  in the dispersionless limit. Dispersionless Lax pairs constitute a key ingredient of the dispersionless $\bar \partial$-method,  and a novel version of the inverse scattering transform  \cite{Bogdanov, Kon, Manakov}.
  It was demonstrated in \cite{Fer5, FKT} that, for a number of particularly interesting classes of systems, the existence of  dispersionless Lax pairs is {\it equivalent} to the existence  of hydrodynamic reductions and, thus, to the integrability. Our third main result (Theorem 3 of Sect. 6) can be formulated as follows:
\begin{itemize}
\item {\bf Any integrable equation  of the form (\ref{u})  possesses a dispersionless
Lax pair. Furthermore, the existence of a dispersionless Lax pair is equivalent to the existence of an infinity of hydrodynamic reductions and, thus, is necessary and sufficient for the integrability.}
\end{itemize}

Differential-geometric aspects of integrable equations of the form (\ref{u}) are discussed in Sect. 7. Our main observation is that the equivalence group $SL(4, R)$ acts by projective transformations on the space of first order derivatives $p^1=u_x, \ p^2=u_y, \ p^3=u_t$, which is thus  identified with the projective space $P^3$, and the coefficient matrix $f_{ij}({\bf p})$ supplies $P^3$ with a well-defined conformal structure
$f_{ij}dp^idp^j$. The classification of integrable equations of the form (\ref{u}) up to the action of the equivalence group $SL(4, R)$ is therefore equivalent to the classification of conformal structures in $P^3$ up to the projective equivalence.  Our first result in this connection is a characterization of linearizable equations (Theorem 4 of Sect. 7.1):
\begin{itemize}
\item {\bf Equation of the form (\ref{u}) is linearizable by a transformation from the equivalence group if and only if the corresponding conformal structure  possesses a quadratic complex of null lines with the Segre symbol [(222)].}
\end{itemize}
A tensorial characterization of linearizable and Lagrangian equations is provided in Sect. 7.2. The integrability conditions impose strong restrictions on the conformal structure $f_{ij}dp^idp^j$ (Theorem 5 of Sect. 7.4):
\begin{itemize}
\item {\bf Conformal structures associated with  integrable equations of the form (\ref{u}) are conformally flat.}
\end{itemize}
Thus, the theory of integrable equations of the form (\ref{u}) has two `flat'  counterparts:
 one is the standard  projective space $P^3$ with coordinates $p^1, p^2, p^3$, and the projective action of $SL(4, R)$;  alternatively, one can speak of a projectively flat connection. Another is a flat conformal structure  $f_{ij}dp^idp^j$. Although, viewed separately, both structures are trivial, this is no longer true when they are imposed simultaneously: their `flat coordinate systems' may not coincide (in fact, they do coincide for linearizable equations only). Our final result provides a differential-geometric characterization of the integrability conditions (Sect. 7.4):
\begin{itemize}
\item {\bf Integrable equations of the form (\ref{u}) correspond to conformal structures in  $P^3$ which possess  an infinity of three-conjugate null coordinate systems   parametrized by three arbitrary functions of one variable.}
\end{itemize}
Notice that we are in the realm of two different geometries, namely, conformal geometry (responsible for the property of being null), and projective geometry (responsible for the property of  being conjugate). It is quite remarkable that the moduli space of such structures is only $20$-dimensional!

\section{Derivation of the integrability conditions: the method of hydrodynamic reductions}

As proposed in \cite{Fer4}, the method of hydrodynamic reductions applies to quasilinear equations of the following general form:
\begin{equation}
A ({\bf u}){\bf u}_x+B({\bf u}){\bf u}_y+C({\bf u}){\bf u}_t=0;
\label{quasi}
\end{equation}
here ${\bf u}=(u^1, ..., u^m)^t$ is an $m$-component column vector of the dependent variables, and $A, B, C$ are $l\times m$ matrices where $l$, the number of equations, is allowed to exceed the number of the unknowns, $m$. The method of hydrodynamic reductions consists of seeking multi-phase solutions in the form
\begin{equation}
{\bf u}={\bf u}(R^1, ..., R^N)
\label{phase}
\end{equation}
where the `phases' $R^i(x, y, t)$ are required to satisfy a pair of consistent equations of hydrodynamic type,
\begin{equation}
 R^i_y=\mu^i(R) R^i_x,  ~~~~ R^i_t=\lambda^i(R) R^i_x.
\label{R}
\end{equation}
We recall that the consistency
(or commutativity) conditions, $R^i_{yt}=R^i_{ty}$,
imply the following restrictions for the
characteristic speeds $\mu^i$ and $\lambda^i$:
\begin{equation}
\frac{\partial_j\mu^i}{\mu^j-\mu^i}=\frac{\partial_j\lambda
^i}{\lambda^j-\lambda^i},
\label{comm}
\end{equation}
$i\ne j, ~ \partial_i=\partial/\partial_{R^i}$, see \cite{Tsarev}.

\medskip

\noindent {\bf Definition. }  \cite{Fer4}
{\it A system (\ref{quasi}) is said to be {\it integrable} if, for any number of phases $N$,  it possesses infinitely many $N$-phase solutions  parametrized by $2N$ arbitrary functions of one variable}.

\medskip

 Integrable equations of the form (\ref{quasi}) arise in a whole range of physical and differential-geometric applications, and contain many particularly important classes of systems. Thus, in the case of square matrices, $m=l$, one arrives at systems of hydrodynamic type,
 $$
 {\bf u}_t+A ({\bf u}){\bf u}_x+B({\bf u}){\bf u}_y=0;
 $$
we recall that $n$-phase solutions for hydrodynamic type systems, also known as solutions with a degenerate hodograph (when $n=1$ or $n=2$), have been extensively investigated in gas dynamics \cite{Sidorov}.
Second order quasilinear PDEs of the form (\ref{u})
can be cast into the form (\ref{quasi}) by setting ${\bf u}=(u_x, u_y, u_t)$;  this gives a system of four equations  in the three unknowns:  $m=3, \ l=4$.
Another interesting subclass  corresponds to equations of the dispersionless Hirota type,
 $$
F(u_{xx}, u_{xy}, u_{yy}, u_{xt}, u_{yt}, u_{tt})=0,
$$
see  \cite{Lenos} and references therein: one can choose any five  second order partial derivatives of the function $u(x, y, t)$ as the dependent variables ${\bf u}$. This corresponds to the  case $m=5,  \ l=8$.
Let us illustrate the method of hydrodynamic reductions using  the dKP equation as an example.

\medskip

\noindent {\bf Example.} Rewriting the dKP equation,
$$
u_{xt}-u_xu_{xx}-u_{yy}=0,
$$
in the new variables $a=u_x, \ b=u_y, \ c=u_t$, one obtains a quasilinear system of four equations in the three unknowns,
\begin{equation}
a_y=b_x, ~~~ a_t=c_x, ~~~ b_t=c_y, ~~~
a_t-aa_x- b_y=0.
\label{dKP}
\end{equation}
Looking for $N$-phase solutions in the form $a=a(R^1, ..., R^N), \ b=b(R^1, ..., R^N), \ c=c(R^1, ..., R^N)$,
where the phases $R^i$ satisfy Eqs. (\ref{R}), one readily obtains the relations
\begin{equation}
\partial_i b=\mu^i \partial_i a, ~~~ \partial_i c=\lambda^i \partial_i a, ~~~  \lambda^i=a+(\mu^i)^2.
\label{110}
\end{equation}
The compatibility conditions
$\partial_i\partial_jb=\partial_j\partial_ib$ and $ \partial_i\partial_jc=\partial_j\partial_ic$ imply
\begin{equation}
\partial_i\partial_ja=\frac{\partial_j\mu^i}{\mu^j-\mu^i}\partial_ia+\frac{\partial_i\mu^j}{\mu^i-\mu^j}\partial_ja,
\label{11}
\end{equation}
while the commutativity conditions (\ref{comm}) result in
\begin{equation}
\partial_j\mu^i=\frac{\partial_j a}{\mu^j-\mu^i}.
\label{12}
\end{equation}
Ultimately, the substitution of (\ref{12}) into (\ref{11}) implies the
following system for $a(R)$ and $\mu^i(R)$,
\begin{equation}
\partial_j\mu^i=\frac{\partial_j a}{\mu^j-\mu^i}, ~~~
\partial_i\partial_ja=2\frac{\partial_ia\partial_ja}{(\mu^j-\mu^i)^2},
\label{13}
\end{equation}
$i\ne j$, which was first derived in \cite{GibTsa96, GibTsa99} in the
context of hydrodynamic reductions of
Benney's moment equations.
For any solution $\mu^i,  a$ of the system (\ref{13}) one can
reconstruct $\lambda^i$ and $b, c$ by virtue of (\ref{110}).
A remarkable feature of the system (\ref{13}) is its multi-dimensional
consistency:  $\partial_k\partial_j\mu^i=\partial_j\partial_k\mu^i$ and $\partial_k\partial_i\partial_ja=\partial_j\partial_i\partial_ka$. The general solution of the system (\ref{13})
depends, modulo reparametrizations $R^i\to f^i(R^i)$, on $N$
arbitrary functions of a single variable. Taking into account that $N$ extra arbitrary functions come from the general solution of Eqs. (\ref{R}), which can be obtained  by the generalised hodograph method \cite{Tsarev}, one ends up with an infinity of $N$-phase solutions depending on $2N$ arbitrary functions of a single variable. We point out that the compatibility conditions,  $\partial_k\partial_j\mu^i=\partial_j\partial_k\mu^i$ and $\partial_k\partial_i\partial_ja=\partial_j\partial_i\partial_ka$, involve {\it triples} of indices $i\ne j \ne k$ only. Thus, the consistency of the system (\ref{13}) for $N=3$ implies
its consistency for arbitrary $N$. This turns out to be a general phenomenon.

\medskip

The  main result of this section is the following

\begin{theorem} The moduli space of integrable equations of the form (\ref{u}) is $20$-dimensional.
\end{theorem}

\medskip
\centerline{\bf Proof:}
\medskip

\noindent
Here we only sketch the proof: full details are provided in the Appendix.
Our strategy is to derive a set of constraints  which are necessary and sufficient for the existence of an infinity of $n$-phase solutions. These constraints constitute a complicated system of second order PDEs
for the coefficients $f_{ij}$, which is in involution; after that, the calculation of the dimension of the moduli space reduces to a simple parameter count.
 The main steps of the derivation of the integrability conditions can be summarized as follows.
First we introducing the variables $a=u_x, \ b=u_y, \ c=u_t$, which transform Eq. (\ref{u})  into the quasilinear form (\ref{quasi}),
$$
a_y=b_x, ~~~ a_t=c_x, ~~~ b_t=c_y, ~~~
f_{11} a_x+ f_{22} b_y + f_{33} c_t+ 2f_{12} a_y  + 2f_{13} a_t  + 2f_{23} b_t=0.
$$
Following the method of hydrodynamic reductions, we seek multi-phase solutions  in the form
$$
a=a(R^1, \ldots ,
R^N), \  \ b=b(R^1, \ldots , R^N), \ \ c=c(R^1, \ldots , R^N),
$$
 where the  phases
$R^1(x,y,t)$, \ldots , $R^N(x,y,t)$ are {\it arbitrary} solutions of
 a pair of commuting hydrodynamic type flows (\ref{R}).
The substitution of this  ansatz into the above quasilinear system leads to  the equations
$$
\partial_i b=\mu^i\partial_i a, ~~~ \partial_i c =\lambda^i\partial_i a,
$$
along with the dispersion relation
$$
D(\lambda^i, \mu^i) =
f_{11} + f_{22}(\mu^i)^2 + f_{33} (\lambda^i)^2 + 2f_{12}\mu^i + 2f_{13} \lambda^i
 + 2f_{23}\mu^i\lambda^i=0.
$$
The consistency  of the equations for $\partial_ib$ and $\partial_ic$ implies
$$
\partial_i\partial_ja=\frac{\partial_j\lambda
^i}{\lambda^j-\lambda^i}\partial_i
a+\frac{\partial_i\lambda
^j}{\lambda^i-\lambda^j}\partial_j a.
$$
Differentiating the dispersion relation with respect to $R^j, \
j\ne i,$ and keeping in mind Eqs.
 (\ref{comm}), one obtains  explicit expressions for $\partial_j
\lambda^i$ and
$\partial_j \mu^i$ in the form
$$
\partial_j\lambda^i=(\lambda^i-\lambda^j)B_{ij}\partial_ja, ~~~
\partial_j\mu^i=(\mu^i-\mu^j)B_{ij}\partial_ja,
$$
where $B_{ij}$ are certain  {\it rational} expressions in $\lambda^i, \lambda^j, \mu^i, \mu^j$,
whose coefficients depend on $f_{ij}(a,b,c)$ and  first order
derivatives thereof (see the Appendix for explicit formulae). Thus,
$$
\partial_i\partial_ja=-(B_{ij}+B_{ji})\partial_ia\partial_ja.
$$
Calculating the consistency conditions $\partial_j\partial_k\lambda^i=\partial_k\partial_j\lambda^i, ~
\partial_j\partial_k\mu^i=\partial_k\partial_j\mu^i$ and $\partial_i\partial_j\partial_ka=\partial_i\partial_k\partial_ja$, one arrives at $30$ differential relations for the coefficients $f_{ij}$, which are linear in the second order derivatives thereof. These relations are manifestly conformally invariant, and without any loss of generality one can set, say, $f_{22}=1$.
Solving for the second order derivatives of the remaining coefficients  $f_{11}, f_{12}, f_{13}, f_{23}, f_{33}$, one obtains $30$ relations  which can be represented in the symbolic form
\begin{equation}
d^2f_{ij}=\frac{1}{F}R(f_{kl}, df_{kl});
\label{int-cond}
\end{equation}
here $F=det \{f_{ij}\}$, and $R$ is a polynomial expression  quadratic in each of its arguments. In other words, any second order derivative of  any coefficient $f_{ij}$ is a certain explicit expression in terms of $f_{kl}$ and first order derivatives thereof. A direct computation shows that the system (\ref{int-cond}) is in involution: all compatibility conditions
are satisfied identically. Since the values of the five functions $f_{11}, f_{12}, f_{13}, f_{23}, f_{33}$, and first order derivatives thereof, are not restricted by any additional constraints, we obtain a $5+3\cdot 5=20$-dimensional moduli space of integrable equations. This finishes the proof of Theorem 1.

The invariant tensorial formulation of the integrability conditions (\ref{int-cond}) is provided in Sect. 7.4.

\medskip

\noindent {\bf Remark.} Notice that, in two dimensions, any equation of
the form
$$
f_{11}(u_x, u_y)u_{xx}+2f_{12}(u_x, u_y)u_{xy}+f_{22}(u_x, u_y)u_{yy}=0
$$
is automatically integrable. Indeed, in the
new variables $a=u_x, \ b=u_y$
it takes the form of a two-component quasilinear system, $a_y=b_x, \
f_{11}(a, b)a_x+2f_{12}(a, b)a_y+f_{22}(a, b)b_y=0$, which linearises under the hodograph transformation
interchanging dependent and independent variables. This trick, however, does not work in higher dimensions.

\section{The equivalence group}

The results of this section are not restricted to the dimension three, and hold  in any dimension. Let us consider a multi-dimensional second order equation of the form
\begin{equation}
\sum f_{ij} u_{x_i x_j}=0
\label{2nd}
\end{equation}
where $u=u(x_1, ..., x_n)$ is a function of $n$ independent variables, and the coefficients $f_{ij}({\bf p})=f_{ij}(p^1, ..., p^n)$ depend on the first order derivatives $p^i=u_{x_i}$ only.
In matrix form,  Eq. (\ref{2nd}) can be represented as
$$
tr FU =0,
$$
where $F=(f_{ij})$ is the (symmetric) matrix of coefficients, defined up to a scalar multiple, and $U$ is the Hessian matrix of the function $u$. Our main observation is that the space $P^n$ with coordinates $p^1, ..., p^n$ admits a natural projective action of the group $SL(n+1, R)$, and  the matrix $f_{ij}$ endows            $P^n$  with a well-defined non-degenerate  conformal structure $f_{ij}dp^idp^j$. This can be seen as follows.
The class of equations (\ref{2nd}) is invariant under linear point transformations of the form
\begin{equation}
\tilde {\bf x}=C{\bf x}+bu, ~~~ \tilde u=c{\bf x}+\beta u,
\label{xu}
\end{equation}
where ${\bf x}=(x_1, ..., x_n)^t$ is a (column) vector of the  independent variables, $C$ is  a $n\times n$ matrix, $b$ and $c$ are $n$-component column and row vectors, respectively, and $\beta$ is a constant. The requirement that the  transformation (\ref{xu}) belongs to the special linear group $SL(n+1, R)$ is equivalent to the condition $det C(\beta-cC^{-1}b)=1$. The induced transformation law for the (row) vector of first order derivatives  ${\bf p}=(p^1, ..., p^n)$ is manifestly projective,
$$
{\bf p}=(\tilde {\bf p}C-c)/(\beta -\tilde {\bf p}b),
$$
with the inverse transformation given by
\begin{equation}
\tilde {\bf p}=\kappa\ {\bf p}C^{-1}+cC^{-1};
\label{tp}
\end{equation}
here the scalar $\kappa$ is defined as $\kappa=\frac{\beta-cC^{-1}b}{1+{\bf p}C^{-1}b}=\frac{1}{det (C+b{\bf p})}$.  A direct calculation gives
\begin{equation}
d\tilde {\bf p}=\kappa d{\bf p} \left( C^{-1}-\frac{C^{-1}b{\bf p}C^{-1}}{1+{\bf p}C^{-1}b}\right)=\kappa d{\bf p}(C+b{\bf p})^{-1}.
\label{dtp}
\end{equation}
Note also that $d\tilde {\bf x}=(C+b{\bf p}) d{\bf x}$.
The transformation law for the Hessian matrix $U$ can be obtained by substituting $d\tilde {\bf p}, \ d\tilde {\bf x}$ into the equality $d\tilde {\bf p}=d\tilde {\bf x}^t \tilde U$:
$$
U=\frac{1}{\kappa} (C+b{\bf p})^t\tilde U(C+b{\bf p}).
$$
Using this expression for $U$ one obtains
$$
tr FU =\frac{1}{\kappa} tr [F (C+b{\bf p})^t\tilde U (C+b{\bf p})]=\frac{1}{\kappa} tr [(C+b{\bf p}) F (C+b{\bf p})^t\tilde U]=
tr \tilde F \tilde U
$$
where the new   coefficient matrix $\tilde F$ is given by
\begin{equation}
\tilde F=\frac{1}{\kappa}(C+b{\bf p})F(C+b{\bf p})^t.
\label{tA}
\end{equation}
Using the formulae (\ref{dtp}) and (\ref{tA})
one readily verifies the identity
$$
d\tilde {\bf p} \tilde F d \tilde {\bf p}^t=\kappa d{\bf p}  F d{\bf p}^t,
$$
which shows that the conformal class of the quadratic form $d{\bf p}  F d{\bf p}^t=f_{ij}dp^idp^j$ is defined in an invariant way. Thus, the coefficient matrix $F({\bf p})$ can be viewed as defining a conformal structure  in the projective space $P^n$ with coordinates $p^1, ..., p^n$ and the standard projective action of  $SL(n+1, R)$. Thus,

\medskip

\noindent
 {\it the study of equations of the form (\ref{2nd}) up to linear transformations of the dependent and independent variables is equivalent to the study of conformal structures in $P^n$ up to the projective equivalence. The global counterpart of Eq. (\ref{2nd}) would be a conformal structure on a projective manifold $M^n$ (a manifold is said to be projective, or endowed with a flat projective structure, if it possesses an atlas with projective transition maps).}

 \medskip

 \noindent
The integrability conditions constitute a  system  of second order PDEs for the coefficient matrix $F({\bf p})$. Since point transformations preserve the integrability, the group generated by Eqs. (\ref{tp}) and (\ref{tA})
constitutes a point symmetry group of the integrability conditions. This $SL(n+1, R)$-invariance plays a crucial role in the further analysis. In particular, all classification results presented below are formulated modulo this equivalence: two
$SL(n+1, R)$-related equations are regarded as the `same'.

Returning to the case $n=3$, we have a $20$-dimensional moduli space of integrable equations, with the action of the equivalence group $SL(4, R)$. One can prove that this action is locally free, that is, its  generic orbits are $15$-dimensional. Thus, up to the action of $SL(4, R)$, a generic  integrable equation is expected to depend on $5$ essential parameters.

\section{Examples and classification results}

Here we present some further examples (both known and new) and partial classification results of integrable PDEs of the form
(\ref{u}) based on the integrability conditions derived in Sect. 2. Although these conditions are quite complicated in general, they can be written down and  solved explicitly under various simplifying assumptions.
A computer  program which calculates the integrability conditions can be obtained from

\noindent {\tt http://www-staff.lboro.ac.uk/\~{}maevf/bft-supplementary-materials-2008.tar.gz}

\noindent
 We construct a whole variety of new integrable examples expressible in elementary functions, elliptic functions and modular forms, thereby manifesting the richness of the problem.

\subsection{List of known examples}

In this subsection we bring together  some of the known examples  which arise, primarily,  in the context of  the dispersionless KP/Toda hierarchies. The simplest generalizations of the dKP equation are
 the modified dKP equation,
$$
u_{xt}+\frac{1}{2}u_x^2u_{xx}-u_{yy}-u_yu_{xx}=0,
$$
and the deformed dKP equation,
$$
u_{xt}+\frac{\epsilon^2}{2} u_x^2u_{xx}-\epsilon u_yu_{xx}-u_{yy}-u_xu_{xx}=0,
$$
see e.g. \cite{Kup1}. Further examples include  dispersionless limits of the KP and Toda  singular manifold equations,
$$
\left(\frac{u_t}{u_x}-\frac{3}{8}\left(\frac{u_y}{u_x}\right)^2\right)_x=\frac{3}{4}\left(\frac{u_y}{u_x}\right)_y
$$
and
\begin{equation}
u_{xy}-\frac{u_xu_y}{(e^{u_t}-1)(1-e^{-u_t})} u_{tt}=0,
\label{dToda}
\end{equation}
respectively \cite{Bogdanov1}. Various symmetric forms  of the dispersionless KP, BKP and Toda  hierarchies were obtained in \cite{Bogdanov1, Bogdanov2}:
\begin{equation}
\begin{array}{c}
(u_y-u_z)u_{yz}+(u_z-u_x)u_{xz}+(u_x-u_y)u_{xy}=0,\\
\ \\
u_x(u_y-u_z)u_{yz}+u_y(u_z-u_x)u_{xz}+u_z(u_x-u_y)u_{xy}=0,\\
\ \\
u_x(u_y^2-u_z^2)u_{yz}+u_y(u_z^2-u_x^2)u_{xz}+u_z(u_x^2-u_y^2)u_{xy}=0,\\
\ \\
(e^{u_x}-1)(e^{u_y}-e^{u_z})u_{yz}+(e^{u_y}-1)(e^{u_z}-e^{u_x})u_{xz}+(e^{u_z}-1)(e^{u_x}-e^{u_y})u_{xy}=0.
\end{array}
\label{A}
\end{equation}
A modification of the standard R-matrix scheme leads to the so-called $r$-th dispersionless modified KP and  $r$-Dym equations,
$$
u_{xy}-\frac{3-r}{(2-r)^2} u_{tt}+\frac{(3-r)(1-r)}{2-r}u_tu_{xx}+\frac{(3-r)r}{2-r}u_xu_{xt}+
\frac{(3-r)(1-r)}{2}u_x^2u_{xx}=0
$$
and
 $$
u_{yt}-\frac{3-r}{2-r}\left(\frac{1}{2-r}u_yu_{xx}-\frac{1}{1-r}u_xu_{xy}\right)=0,
$$
see \cite{Bla}, \cite{Bla1}, \cite{Manas}. Further examples arise in the context of  the so-called `universal hierarchy':
$$
u_{yy}=u_yu_{xz}-u_{xy}u_z,  ~~~~
u_{xz}=u_xu_{yy}-u_{xy}u_y,  ~~~~
\left(\frac{u_t}{u_x}\right)_t=\left(\frac{u_y}{u_x}\right)_x,
$$
see \cite{Shabat, Ovsienko}.  A  class of integrable Euler-Lagrange equations of the form
$$
(g_{u_x})_x+(g_{u_y})_y+(g_{u_t})_t=0
$$
was investigated in \cite{FKT}; here the Lagrangian density $g$ is a function of the first order derivatives only,
$g=g(u_x, u_y, u_t)$. Among the apparently new examples found in \cite{FKT} one should mention  the equation
$$
u_tu_{xy}+u_xu_{yt}+u_yu_{xt}=0,
$$
which corresponds to the Lagrangian density $g=u_xu_yu_t$. It was demonstrated in \cite{Odesskii} that the generic integrable Lagrangian density $g$ is a modular form of its arguments.

Further examples of integrable equations of the form (\ref{u}) result,
 via appropriate substitutions, from   $2$D integrable systems of hydrodynamic type. For instance, the paper \cite{Moro} provides a complete list of integrable Hamiltonian systems of the form
\begin{equation}
v_t+(H_v)_y=0, ~~~ w_x+(H_w)_y=0
\label{Ham}
\end{equation}
where the Hamiltonian density $H$ is a function of $v, w$. Eq. (\ref{Ham})$_2$ implies the existence of a potential $u$ such that $u_y=w, \ u_x=-H_w$. Expressing
$v$ and $w$ in terms of $u_x$ and $u_y$ and substituting these expressions into Eq. (\ref{Ham})$_1$, one obtains an equation of the form (\ref{u}). Thus, the simplest integrable example $H=\frac{1}{2}vw^2$ results in the equation $u_xu_{yt}-u_yu_{xt}+u_y^3u_{yy}=0$, etc.

Another possible construction exploits the fact that any two-component integrable system of hydrodynamic type  possesses a dispersionless Lax pair of the form
$$
S_t=F(S_x, v, w), ~~ S_y=G(S_x, v, w),
$$
see \cite{Fer5}. Here  $v$ and $w$ are the dependent variables which satisfy a system of hydrodynamic type resulting from the compatibility conditions $S_{yt}=S_{ty}$. Expressing $v$ and $w$ in terms of  $S_x, S_y$ and $S_t$, and substituting these expressions into the equations for $v$ and $w$, one obtains a single second order equation of the form (\ref{u}) for  $S$. This construction is analogous to the  `eigenfunction equations'  in soliton theory \cite{Ko}.

We have verified that all examples listed in this section indeed satisfy the integrability conditions of Sect. 2.

\subsection{Equations of the form $u_{tt} = f(u_x, u_y, u_t)(u_{xx}+u_{yy})$}

For these  equations, which can be viewed as nonlinear analogues of the $(2+1)$-dimensional wave equation, the integrability conditions take the form
\begin{eqnarray*}\label{utt=f(uxx+uyy)}
    2ff_{bb} + f_a^2 - 3f_{b}^2 &= 0, \notag\\
    2ff_{aa} - 3f_a^2 + f_{b}^2 &= 0, \notag \\
    2f^2f_{cc} - 2ff_{c}^2 + f_{a}^2 + f_{b}^2 &= 0,\\
    f_{a}f_{c} - f_{ac}f &= 0, \notag \\
    2f_{a}f_{b} - f_{ab}f &= 0,\notag \\
    f_{b}f_{c} - f_{bc}f &= 0,\notag
\end{eqnarray*}
recall that $a = u_x,\ b = u_y,\ c = u_t$. These relations are straightforward to solve. The case $f_a=f_b=0$ leads, up to equivalence transformations, to a unique solution $f=e^ c$, which corresponds to  the Boyer-Finley equation
$$
u_{tt} = e^{u_t}(u_{xx}+u_{yy}).
$$
Assuming $f_a$ and $f_b$ to be nonzero (notice that $f_a$ and $f_b$ can only be zero or nonzero simultaneously),  the last three equations imply
\begin{equation*}
    f(a,b,c) =  \frac{p(c)}{\varphi(a)+\psi(b)},
\end{equation*}
and the substitution into the remaining equations gives
\begin{equation*}
    f(a,b,c) = \frac{\alpha e^{\beta c} + \gamma e^{-\beta c} + \delta}{a^2+b^2+\xi},
\end{equation*}
where the constants $\alpha, \beta, \gamma, \delta, \xi$ satisfy a single quadratic constraint
    $4\alpha\gamma - \delta^2 = 0$. This corresponds to equations of the form
$$
u_{tt} = \frac{\alpha e^{\beta u_t} + \gamma e^{-\beta u_t} + \delta}{u_x^2+u_y^2+\xi}(u_{xx}+u_{yy}),
$$
which are analogous to the singular manifold dToda equation (\ref{dToda}).

\subsection{Equations of the form $u_{tt}=f(u_x, u_y, u_t)_x$}

These equations  are related to the Hirota-type equations
$\Omega_{tt}=f(\Omega_{xx}, \Omega_{xy}, \Omega_{xt})$ via a substitution $u=\Omega_x$. In this form they were investigated in
\cite{Pavlov1, FKP}.
The integrability conditions take the form
\begin{eqnarray*}
&&f_{ccc}=\frac{2f_{cc}^{2}}{f_{c}},\quad
f_{acc}=\frac{2f_{ac}f_{cc}}{f_{c}}, \quad
f_{bcc}=\frac{2f_{bc}f_{cc}}{f_{c}},  \notag \\
&&f_{aac}=\frac{2f_{ac}^{2}}{f_{c}}, \quad
f_{abc}=\frac{2f_{ac}f_{bc}}{f_{c}}, \quad
f_{bbc}=\frac{2f_{bc}^{2}}{f_{c}},  \notag \\
&&f_{bbb}=\frac{2}{f_{c}^{2}}\left(
f_{b}f_{bc}^{2}+f_{bc}(f_{c}f_{bb}+2f_{ac})-f_{cc}(f_{b}f_{bb}+2f_{ab})
\right),  \notag \\
&&f_{abb}=\frac{2}{f_{c}^{2}}\left(
f_{a}f_{bc}^{2}+f_{ac}(f_{c}f_{bb}+f_{ac})-f_{cc}(f_{a}f_{bb}+f_{aa})\right), \\
&&f_{aab}=\frac{2}{f_{c}^{2}}\left(
f_{cc}(f_{b}f_{aa}-2f_{a}f_{ab})-f_{ac}(f_{b}f_{ac}-2f_{c}f_{ab})-f_{bc}(f_{c}f_{aa}-2f_{a}f_{ac})\right),
\notag \\
&&f_{aaa}=\frac{2}{f_{c}^{2}}\left(
(f_{a}+f_{b}^{2})f_{ac}^{2}+f_{a}^{2}f_{bc}^{2}+f_{c}^{2}(f_{ab}^{2}-f_{aa}f_{bb})-f_{bb}f_{cc}f_a^2 \right.
\notag \\
&&\qquad
+f_{ac}f_{c}(f_{aa}+2(f_{a}f_{bb}-f_{b}f_{ab}))+2f_{bc}(f_{b}(f_{c}f_{aa}-f_{a}f_{ac})-f_{a}f_{c}f_{ab})
\notag \\
&&\qquad -\left.
f_{cc}((f_{a}+f_{b}^{2})f_{aa}-2f_{a}f_{b}f_{ab})\right). \notag
\end{eqnarray*}
This system is in involution, and its general solution depends on $10 $
arbitrary constants. The integration  leads to the four essentially different canonical forms,
\begin{eqnarray*}
f &=&u_{y}+\frac{1}{4A}(Au_{t}+2Bu_{x})^{2}+Ce^{-Au_{x}}, \\
f &=&\frac{u_{y}}{
u_{x}}+\left( \frac{1}{u_{x}}
+\frac{A}{4u_{x}^{2}}\right) u_{t}^{2}+\frac{B}{u_{x}^{2}}u_{t}+\frac{B^{2}}{Au_{x}^{2}}+Ce^{A/u_{x}},
\\
f &=&\frac{u_{y}}{u_{t}}+\frac{1}{6}\eta (u_{x})u_{t}^{2}, \\
f &=&\ln u_{y}-\ln \theta _{1}\left(u_{t}, u_{x}\right) -\frac{1}{4}\overset{u_{x}}{\int }\eta (\tau )d\tau,
\end{eqnarray*}
see \cite{Pavlov1}. Here $A, B, C$ are arbitrary constants,  $\eta $ is a solution of the Chazy equation  \cite{Chazy},
$$
\eta^{\prime \prime \prime }+2\eta \eta ^{\prime \prime }=3(\eta^{\prime}) ^2,
$$
and $\theta_1$ is the Jacobi  theta-function.

\subsection{Equations of the form $f(u_x, u_y, u_t) u_{xy}+g(u_x, u_y, u_t) u_{xt}+h(u_x, u_y, u_t) u_{yt}=0$}

Equations of this type arise in the context of the dispersionless KP, BKP and Toda hierarchies, see (\ref{A}). Further examples were constructed  in \cite{Adler} in the search for consistent triples of second order equations.
First of all, the integrability conditions imply that any  equation within this class can be reduced to a simplified form,
$$
f(u_x, u_y) u_{xy}+g(u_x,  u_t) u_{xt}+h(u_y, u_t) u_{yt}=0,
$$
via  a multiplication  by an appropriate scalar factor. In terms of the coefficients $f(a, b),  g(a, c)$ and $ h(b, c)$, the integrability conditions  take the form

\begin{align}
 f_{ab} &=  \displaystyle    \frac{h_{b}}{h}f_a
          +  \frac{g_{a}}{g} f_{b} -  \frac{g_{a}}{g} \frac{h_{b} }{h}f,\notag\\
 g_{ac} &= \displaystyle    \frac{h_{c}}{h}g_a
          +  \frac{f_{a}}{f} g_{c} -  \frac{f_{a}}{f} \frac{h_{c} }{h}g,\notag\\
 h_{bc} &= \displaystyle    \frac{g_{c}}{g}h_b
          +  \frac{f_{b}}{f} h_{c} -  \frac{f_{b}}{f} \frac{g_{c} }{g}h,\notag\\
    f_{aa} & =
         \frac{gf_{a} + fg_{a}}{fg} f_{a}
       + \frac{fg_{a} - gf_{a}}{fh} f_{b}
       + \frac{gh_{b} - f h_{c}}{h^2} f_{a}
       + \frac{fg_{a}h_{c} - gg_{a}h_{b}}{gh^2} f,\notag\\
    f_{bb} & =
         \frac{fh_{b} + hf_{b}}{fh} f_{b}
       + \frac{fh_{b} - hf_{b}}{fg} f_{a}
       + \frac{hg_{a} - f g_{c}}{g^2} f_{b}
       + \frac{f h_{b}g_{c} - h g_{a}h_{b}}{g^2h} f, \notag\\
    g_{cc} & =
         \frac{gh_{c} + hg_{c}}{gh} g_{c}
       + \frac{gh_{c} - hg_{c}}{fg} g_{a}
       + \frac{hf_{a} - g f_{b}}{f^2} g_{c}
       + \frac{gf_{b}h_{c} - h f_{a}h_{c}}{f^2h} g,\notag\\
    g_{aa} & =
         \frac{fg_{a} + gf_{a}}{fg} g_{a}
       + \frac{gf_{a} - fg_{a}}{gh} g_{c}
       + \frac{fh_{c} - g h_{b}}{h^2} g_{a}
       + \frac{gf_{a}h_{b} - f f_{a}h_{c}}{fh^2} g,\notag\\
    h_{cc} & =
         \frac{gh_{c} + hg_{c}}{gh} h_{c}
       + \frac{hg_{c} - gh_{c}}{fh} h_{b}
       + \frac{gf_{b} - h f_{a}}{f^2} h_{c}
       + \frac{hf_{a}g_{c} - g f_{b}g_{c}}{f^2g} h,\notag\\
    h_{bb} & =
         \frac{gh_{b} + hf_{b}}{fh} h_{b}
       + \frac{hf_{b} - fh_{b}}{gh} h_{c}
       + \frac{fg_{c} - h g_{a}}{g^2} h_{b}
       + \frac{hg_{a}f_{b} - f f_{b}g_{c}}{fg^2} h. \notag
\end{align}

 \medskip

\noindent Remarkably, the first three equations appeared previously in \cite{Moro} in the problem of  classification of integrable Hamiltonian systems of hydrodynamic type. It was observed that their general solution is representable in the form
\begin{equation}
f=\frac{p(a)-q(b)}{P(a)Q(b)}, ~~~
g=\frac{r(c)-p(a)}{P(a)R(c)}, ~~~
h=\frac{q(b)-r(c)}{Q(b)R(c)},
\label{fgh}
\end{equation}
where $p, q, r$ and $P, Q, R$ are arbitrary functions of the indicated arguments. The further analysis splits into four different cases depending on how many functions among $p, q, r$ are constant.
In the simplest case when all three of them are constant, the integrability conditions reduce to $P''=Q''=R''=0$. Thus, any equation of the form
$$
   a_1 u_x u_{yt} + a_2 u_y u_{xt} + a_3 u_t u_{xy} = 0
$$
is automatically integrable. Let us concentrate on the generic case when none of $p, q, r$ are constant (intermediate cases  can be considered in a similar way).
Under the substitution (\ref{fgh}), the first three integrability conditions  will be satisfied identically, while the last six imply a system of second order ODEs for  $p, q, r$ and $P, Q, R$:

\begin{align}
    \frac{P''}{p'} &= \frac{P'-Q'}{p-q} + \frac{P'-R'}{p-r}- \frac{Q'-R'}{q-r} ,\notag\\
    \frac{R''}{r'} &=  \frac{R'-Q'}{r-q} + \frac{R'-P'}{r-p}-\frac{Q'-P'}{q-p},\label{R1}\\
    \frac{Q''}{q'} &= \frac{Q'-P'}{q-p} + \frac{Q'-R'}{q-r} - \frac{P'-R'}{p-r},\notag
    \end{align}
    \begin{align}
    \frac{p''}{p'}P &=  \frac{q-p}{q-r}\left( \frac{Rr'}{r-p} - R' \right) +
            \frac{r-p}{r-q}\left( \frac{Qq'}{q-p} - Q' \right) +
            \frac{p'P (q - 2p + r)}{(p-r)(p-q)},\notag\\
    \frac{r''}{R'}R &=  \frac{q-r}{q-p}\left( \frac{Pp'}{p-r} - P' \right) +
            \frac{p-r}{p-q}\left( \frac{Qq'}{q-r} - Q' \right) +
            \frac{r'R (p - 2r + q)}{(r-p)(r-q)},\label{r}\\
    \frac{q''}{q'}Q &=  \frac{p-q}{p-r}\left( \frac{Rr'}{r-q} - R' \right) +
            \frac{r-q}{r-p}\left( \frac{Pq'}{p-q} - P' \right) +
            \frac{q'Q (p - 2q + r)}{(q-p)(q-r)}.\notag
\end{align}

\noindent The separation of variables in Eqs. (\ref{R1}) implies
$$
P'= F(p), ~~ Q'= F(q), ~~ R'= F(r),
$$
where $F(\cdot)$ is an arbitrary quadratic polynomial, $F(x)= \epsilon x^2+m_1x+m_0$; here the parameters $\epsilon, m_1, m_0$ play the role of separation constants. A similar separation of variables in Eqs. (\ref{r}) results in
$$
p'P= S(p), ~~ q'Q= S(q), ~~ r'R= S(r),
$$
where $S(\cdot)$ is a cubic polynomial, $S(x)=\epsilon x^3+n_2x^2+n_1x+n_0$. These equations lead to $P'/P=p'F(p)/S(p), \ Q'/Q=q'F(q)/S(q), \ R'/R=r'F(r)/S(r)$, and the integration yields
$$
P=c_1(p-\alpha)^{\mu}(p-\beta)^{\nu}(p-\gamma)^{\eta}, ~~
Q=c_2(q-\alpha)^{\mu}(q-\beta)^{\nu}(q-\gamma)^{\eta}, ~~
R=c_3(r-\alpha)^{\mu}(r-\beta)^{\nu}(r-\gamma)^{\eta};
$$
here $c_i$ are three extra integration constants, $\alpha, \beta, \gamma$ are the roots of the polynomial $S$, and the exponents $\mu, \ \nu, \ \eta$, which are related to the coefficients $m_1, m_0$ of the quadratic polynomial $F$,  satisfy a single relation $\mu+\nu+\eta=1$. Thus, setting $\epsilon=1$, we have
$$
p'=\frac{1}{c_1}(p-\alpha)^{1-\mu}(p-\beta)^{1-\nu}(p-\gamma)^{1-\eta}, ~~
q'=\frac{1}{c_2}(q-\alpha)^{1-\mu}(q-\beta)^{1-\nu}(q-\gamma)^{1-\eta},
$$
$$
r'=\frac{1}{c_3}(r-\alpha)^{1-\mu}(r-\beta)^{1-\nu}(r-\gamma)^{1-\eta};
$$
recall that quadratures of this type arise in the context of the Schwarz-Christoffel mappings of triangular domains. Particularly interesting examples correspond to the symmetric choice
$\mu=\nu=\eta=1/3$. In this case the ODEs for $p, q, r$ take the form
\begin{equation}
p'^3=\frac{1}{c_1^3}S^2(p), ~~ q'^3=\frac{1}{c_2^3}S^2(q), ~~ r'^3=\frac{1}{c_3^3}S^2(r).
\label{fgh111}
\end{equation}
We point out that the ansatz (\ref{fgh}) possesses the
obvious $SL(2, R)$-symmetry,
$$
p\to \frac{\alpha p+\beta}{\gamma p+\delta}, ~~ q\to \frac{\alpha q+\beta}{\gamma q+\delta}, ~~
r\to \frac{\alpha r+\beta}{\gamma r+\delta}, ~~~~~
P\to \frac{P}{\gamma p+\delta}, ~~ Q\to \frac{Q}{\gamma q+\delta}, ~~ R\to \frac{R}{\gamma r+\delta},
$$
which can be used to bring the polynomial $S$  to a canonical form. In the case of three distinct roots  one can reduce  $S$  to a quadratic, $S(x)=x^2+g_3$, so that the ODEs (\ref{fgh111}) imply $p=\wp'(a_1u_1), \ q=\wp'(a_2u_2), \ r=\wp'(a_3u_3)$ where  $27a_i^3=2/c_i^3$, and $\wp$ is the Weierstrass $\wp$-function: $(\wp')^2=4\wp^3-g_3$ (notice that we are dealing with an equianharmonic case: $g_2=0$).  This leads  to the integrable equation
$$
\frac{\wp'(a_1u_x)-\wp'(a_2u_y)}{\wp(a_1u_x)\wp(a_2u_y)}\ u_{xy}+
\frac{\wp'(a_3u_t)-\wp'(a_1u_x)}{\wp(a_1u_x)\wp(a_3u_t)}\ u_{xt}+
\frac{\wp'(a_2u_y)-\wp'(a_3u_t)}{\wp(a_2u_y)\wp(a_3u_t)}\ u_{yt}=0.
$$
Up to an appropriate rescaling, this equation is equivalent to
$$
\alpha\frac{\wp'(u_x)-\wp'(u_y)}{\wp(u_x)\wp(u_y)}\ u_{xy}+
\beta\frac{\wp'(u_t)-\wp'(u_x)}{\wp(u_x)\wp(u_t)}\ u_{xt}+
\gamma \frac{\wp'(u_y)-\wp'(u_t)}{\wp(u_y)\wp(u_t)}\ u_{yt}=0.
$$
It possesses a degeneration $g_3\to 0, \ \wp (x)\to 1/x^2$, resulting in
$$
\alpha u_t(u_x^3-u_y^3)u_{xy}+\beta u_y(u_t^3-u_x^3)u_{xt}+\gamma u_x(u_y^3-u_t^3)u_{yt}=0,
$$
compare with (\ref{A}).

\subsection{Integrable equations in terms of  modular forms and theta functions}

This section contains a number of more `exotic'  integrable examples which are not expressible in elementary functions.

Let us begin with  equations of the form
$$
\left( u_y p(u_t) \right)_x + \left( u_x q(u_t) \right)_y + \left( u_x u_y r(u_t) \right)_t = 0.
$$
The integrability conditions yield a complicated system of three third order ODEs  for the functions $p, q$ and $r$. Let us analyze  special cases.

\medskip

\noindent {\bf Case 1.}
The choice $q = p$,  $r = p'$ corresponds to Lagrangian equations with the Lagrangian density $u_xu_yp(u_t)$. In this case the integrability conditions
reduce to a single fourth order ODE for $p$,
\begin{equation*}
    p''''(p^2{p''}-2pp'^2) - p^2{p'''}^2  + 2p{p'}{p''}{p'''}
          + 8{p'}^3{p'''} - 9{p'}^2{p''}^2 = 0.
\end{equation*}
It was shown in \cite{Odesskii} that the generic  solution of this equation  is  a  modular form of level three,  known as the Eisenstein series $E_{1, 3}(z)$:
\begin{equation*}
    p(u_t) = \sum_{(\alpha,\beta)\in Z^2}e^{(\alpha^2-\alpha\beta+\beta^2)u_t}=1 + 6 e^{u_t} + 6 e^{3u_t} + 6e^{4 u_t} + 12 e^{7 u_t} + ...;
\end{equation*}
(the Eisenstein series $E_{1, 3}(z)$ results upon the substitution $u_t\to 2\pi i z$). It can also be written in the form
$$
p(u_t)=1-6\sum_{k=1}^{\infty}\left(\frac{e^{(3k-1)u_t}}{1-e^{(3k-1)u_t}}-\frac{e^{(3k-2)u_t}}{1-e^{(3k-2)u_t}}\right).
$$
Notice that a similar choice $q=p, r=-p'$ corresponds to equations of the form
$u_{xy}-\frac{p''}{2p}u_xu_yu_{tt}=0$. Here the integrability conditions result in a fourth order ODE
$$
p^2p''p''''-p^2p'''^2-2pp''^3+p'^2p''^2=0,
$$
whose properties are quite different from those of the above equation: first of all, one can reduce the order by setting $p''=2ph$. This results in the second order ODE $hh''-h'^2-2h^3=0$, which implies
$$
h(u_t)=4s^2\frac{e^{2su_t}}{(e^{2su_t}-1)^2},
$$
$s={\rm const}$. This is the case of the singular manifold dToda equation (\ref{dToda}), see also Ex. 2 of Sect. 5.

\medskip

\noindent {\bf Case 2.}
Another interesting choice is $p = 1, \  q=0$ which corresponds to equations of the form
$u_{xy}+(u_xu_yr(u_t))_t=0$. The integrability conditions result in a single third order ODE for $r$,
\begin{equation*}
r'''(r'-r^2) -r''^2+ 4{r}^3{r''}+2r'^3 - 6{r}^2{r'}^2= 0,
\end{equation*}
which  appeared recently in a different context  in the theory of  modular forms of level two: set $r=y/2$ to obtain Eq. (4.7) from \cite{A2}. This equation possesses a remarkable $SL(2, R)$-invariance,
$$
\tilde z=\frac{\alpha z+ \beta }{\gamma z+ \delta}, ~~~ \tilde r= (\gamma z+ \delta)^2 r+\gamma(\gamma z+ \delta), ~~~ \alpha \delta-\gamma \beta=1;
$$
here $z=u_t$. Modulo this $SL(2, R)$-action, the generic solution is given by the series
\begin{equation*}
    r(u_t) = 1 + 8 e^{4u_t} - 8 e^{8u_t} + 32 e^{12u_t} -40e^{16u_t}+48e^{20u_t}-32e^{24u_t}+ ...,
\end{equation*}
which, upon setting $e^{4u_t}=q$, coincides with the Eisenstein series,
$$
{\cal E}(q)=1-8\sum_{n=1}^{\infty}\frac{(-1)^nnq^n}{1-q^n},
$$
associated with the congruence subgroup $\Gamma_0(2)$ of the modular group \cite{A2}.

\medskip

\noindent {\bf Case 3.}
As a generalization of Case 2, let us consider equations of the form
$u_{xy} + (u_x r(u_y,u_t))_t = 0$. The integrability conditions take the form
\begin{align*}
 \frac{r_{bbb}}{r_{bb}} &= \frac{rr_{bb}-r_{bc}+r_{b}^2}{rr_{b}-r_{c}},\\
 \frac{r_{bbc}}{r_{bb}} &= \frac{rr_{bc}-r_{cc}+r_{b}r_{c}}{rr_{b}-r_{c}},\\
 r_{bcc} &= \frac{1}{rr_{b}-r_{c}}(2rr_{bc}^2-rr_{bb}r_{cc}-r_{cc}r_{bc}+r_{b}^2r_{cc}-2r_{b}r_{c}r_{bc}+2r_{c}^2r_{bb}),\\
 r_{ccc} &= \frac{1}{rr_{b}-r_{c}}(2r^2r_{bc}^2-2r^2r_{bb}r_{cc}+rr_{cc}r_{bc}+4rr_{b}^2r_{cc}-8rr_{b}r_{c}r_{bc}+\\
        &4rr_{c}^2r_{bb}-r_{cc}^2-r_{b}r_{c}r_{cc}+2r_{c}^2r_{bc}).
\end{align*}
Here the first two equations imply $r_{bb} = \alpha (rr_{b}-r_{c})$, which is the well-known Burgers equation.
Without any loss of generality we will assume  $\alpha = 1$.
Under the substitution $r=2v_b/v$, the Burgers equation linearizes to the heat equation,
$v_c = v_{bb}$. Modulo this equation, the integrability conditions reduce to a single
sixth order ODE for the function $v$,

\begin{align*}
&        v^4 {v'''} {v^{(6)}} - v^4 {v^{(4)}} {v^{(5)}} + 3 v^3 {v''}^2 {v^{(5)}} -
         5 v^3 {v''} {v'''} {v^{(4)}} + 2 v^3 {v'''}^3 - 3 v^3 {v'} {v''} {v^{(6)}}
     - 2 v^3 {v'} {v'''} {v^{(5)}}\\
&        + 5 v^3 {v'} {v^{(4)}}^2 + 2 v^2 {v'} {v''}
         {v'''}^2 + 6 v^2 {v'}^2 {v''} {v^{(5)}} - 10 v^2 {v'}^2 {v'''} {v^{(4)}}
         + 2 v^2 {v'}^3 {v^{(6)}} - 6 v {v'}^2 {v''}^2 {v'''}\\
&    + 12 v {v'}^3 {v'''}^2 - 6 v {v'}^4 {v^{(5)}} + 6 {v'}^3 {v''}^3 - 12 {v'}^4 {v''}
          {v'''} + 6 {v'}^5 {v^{(4)}} = 0,
\end{align*}
here prime denotes differentiation with respect to $b$. One can show that the generic solution of the
heat equation constrained  by this sixth order ODE is given by the formula
$ v(u_y,u_t) = \theta\left(\frac{u_y}{2\pi},-\frac{u_t}{\pi i} \right),$
where $\theta$ is the Jacobi theta-function:
\begin{equation*} \theta(z,\tau) = 1 + 2\sum_{n=1}^\infty e^{\pi i n^2 \tau} \cos(2\pi n z).
\end{equation*}

\noindent {\bf Remark.} Further generalization,  $u_{xy}+f(u_x, u_y, u_t)_t=0$, was discussed in \cite{FKP} in the dispersionless Hirota form $\Omega_{xy}+f(\Omega_{xt}, \Omega_{yt}, \Omega_{tt})=0$ (these two representations are related via $u=\Omega_t$). It was demonstrated that the generic solution is given by a ratio of two Jacobi theta functions:
$$
f(u_x, u_y, u_t)=-\frac{1}{4}\ln \frac{\theta_1(u_t, u_y-u_x)}{\theta_1(u_t, u_y+u_x)}.
$$

\section{Conservation laws}

Let us begin with some general remarks which clarify the geometric meaning of conservation laws of Eq. (\ref{2nd}). Consider  a first order conservation law in the form
\begin{equation}
\partial_{x_1} g_1+ ... + \partial_{x_n} g_n=0,
\label{con}
\end{equation}
where $g_i({\bf p})$ are functions of the first order derivatives $p^1=u_{x_1}, ..., p^n=u_{x_n}$. The requirement that the equality (\ref{con}) is satisfied identically modulo Eq. (\ref{2nd}) is equivalent to the set of relations $\partial_{p^i}g_i=kf_{ii}, \ \partial_{p^i}g_j+\partial_{p^j}g_i=2kf_{ij}$, here k is a proportionality factor. These  relations can be rewritten in the form
\begin{equation}
dg_1 dp^1+... + dg_n dp^n = k f_{ij} dp^i dp^j,
\label{emb}
\end{equation}
which has a clear geometric interpretation. Let us introduce a fiber bundle $B$ with the base $P^n$ supplied with  a flat conformal structure $dg_idp^i$ (here $p^i$ are coordinates on the base $P^n$,  $g_i$ are coordinates along the fiber). The identity (\ref{emb})  provides an embedding of the conformal structure $f_{ij}dp^idp^j$, defined on the base, into the fiber bundle $B$  with the flat conformal structure $dg_idp^i$. To uncover the geometry of $B$ let us return to the equivalence group $SL(n+1, R)$ which acts via (\ref{xu}) on the space of independent variables $x_i$. The induced action on  conservation laws ${\bf g}=(g_1, ..., g_n)^t$ is
$$
\tilde {\bf g}=\kappa (C+b{\bf p}) {\bf g},
$$
recall that $\kappa =\frac{1}{det (C+b{\bf p})}$, see Sect. 3. Combining this with the transformation law
$$
d\tilde {\bf p}=\kappa d{\bf p}(C+b{\bf p})^{-1}
$$
we obtain
\begin{equation}
d\tilde {\bf p}\  \tilde {\bf g}=\kappa^2 d {\bf p}\   {\bf g}.
\label{kappa}
\end{equation}
Introducing $K=dp^1\wedge ...\wedge dp^n$ we can see that $\tilde K=\kappa ^{n+1}K$ so that (\ref{kappa}) takes the form
\begin{equation}
d\tilde {\bf p}\  \tilde {\bf g}\ {\tilde K}^{-\frac{2}{n+1}}= d {\bf p} \  {\bf g}\ {K}^{-\frac{2}{n+1}}.
\label{pg}
\end{equation}
This shows that ${\bf g}$ has a natural interpretation as a $1$-form with values in the canonical bundle
$K^{-\frac{2}{n+1}}$, namely, ${\bf g}\in H^0(P^n, \ \Omega^1\otimes K^{-\frac{2}{n+1}})$. Notice that for $n=3$ (which is our main case of interest) this bundle appeared in \cite{Hitchin} in the geometric theory of solutions of Einstein's equations.

\medskip

Let us return to the $3$-dimensional case. Our main result is that any integrable equation of the form (\ref{u}) possesses exactly four first order  conservation laws of the form  (\ref{conserv-law}),
\begin{equation*}
    g_1(u_x,u_y,u_t)_x + g_2(u_x,u_y,u_t)_y + g_3(u_x,u_y,u_t)_t = 0.
\end{equation*}
According to the discussion above, this means that there exists a 4-parameter linear family of sections ${\bf g}
\in H^0(P^3, \ \Omega^1\otimes K^{-\frac{1}{2}})$   providing conformal embeddings of the conformal structure $f_{ij}dp^idp^j$. Let us begin with illustrating examples.

\medskip

\noindent {\bf Example 1.} The equation
\begin{equation*}
   a_1 u_x u_{yt} + a_2 u_y u_{xt} + a_3 u_t u_{xy} = 0,
\end{equation*}
which was found in Sect. 4.4,  possesses four conservation  laws of the form
\begin{equation*}
\begin{array}{c}
   (a_2+a_3- a_1) (u_yu_t)_x + (a_1+a_3-a_2)(u_xu_t)_y + (a_1+a_2-a_3) (u_xu_y)_t = 0, \\
    \ \\
    a_2 \left( {u_t}^{\frac{a_1+a_2}{a_3}}u_y\right)_x  + a_1 \left(  {u_t}^{\frac{a_1+a_2}{a_3}}u_x\right)_y = 0, \\
    \ \\
  a_3 \left( {u_y}^{\frac{a_1+a_3}{a_2}}u_t\right)_x  + a_1 \left(  {u_y}^{\frac{a_1+a_3}{a_2}}u_x\right)_t = 0, \\
   \ \\
    a_3 \left( {u_x}^{\frac{a_2+a_3}{a_1}}u_t\right)_y  + a_2 \left(  {u_x}^{\frac{a_2+a_3}{a_1}}u_y\right)_t = 0.
\end{array}
\end{equation*}

\medskip



\noindent {\bf Example 2.} The  dToda singular manifold equation \cite{Bogdanov1},
$$
u_{xy}-\frac{u_xu_y}{(e^{u_t}-1)(1-e^{-u_t})} u_{tt}=0,
$$
possesses two conservation laws of the form
$$
    \left( \log\frac{u_y}{e^{u_t}-1}\right)_x + \left( \frac{u_x e^{u_t}}{e^{u_t}-1}\right)_t = 0, ~~~
    \left( \log\frac{u_x}{e^{u_t}-1}\right)_y + \left( \frac{u_y e^{u_t}}{e^{u_t}-1}\right)_t = 0,
$$
as well as
$$
    \left(p(u_t)u_y\right)_x+    \left(p(u_t)u_x\right)_y - \left(p'(u_t)u_xu_y\right)_t = 0
$$
where   $p$  satisfies the second order linear ODE
\begin{align*}
    \frac{p''}{2p} = \frac{e^{u_t}}{(e^{u_t}-1)^2};
\end{align*}
as pointed out by M. Pavlov, it possesses two linearly independent solutions,
$$
p=\frac{e^{u_t}+1}{e^{u_t}-1} ~~~ {\rm and} ~~~ p=\frac{e^{u_t}+1}{e^{u_t}-1}u_t -2,
$$
which provide two extra conservation laws.


\medskip

\noindent {\bf Example 3.} The  equation
$$
u_t(u_x^3-u_y^3)u_{xy}+ u_y(u_t^3-u_x^3)u_{xt}+ u_x(u_y^3-u_t^3)u_{yt}=0,
$$
which appeared in Sect. 4.4,
possesses four conservation  laws,
$$
(u_xu_y(u_x^3-u_y^3))_t+(u_xu_t(u_t^3-u_x^3))_y+(u_yu_t(u_y^3-u_t^3))_x=0,
$$
as well as
$$
\left[G\left(\frac{u_t}{u_y}\right)\right]_x-\left[G\left(\frac{u_x}{u_y}\right)\right]_t=0, ~~~
\left[G\left(\frac{u_y}{u_x}\right)\right]_t-\left[G\left(\frac{u_t}{u_x}\right)\right]_y=0, ~~~
\left[G\left(\frac{u_x}{u_t}\right)\right]_y-\left[G\left(\frac{u_y}{u_t}\right)\right]_x=0,
$$
here the function $G(s)$ is defined as $G'=\frac{1}{s^3-1}$. Explicitly, one has
$$
G(s)=\frac{1}{3}\left(\ln (s-1)+\epsilon \ln (s-\epsilon)+\epsilon^2 \ln (s-\epsilon^2)\right),
$$
here $\epsilon=e^{2\pi i/3}, ~~ \epsilon^3=1$; notice that  $G$ is real-valued.

\medskip

\noindent {\bf Example 4.} The  equation
$$
\frac{\wp'(u_x)-\wp'(u_y)}{\wp(u_x)\wp(u_y)}\ u_{xy}+
\frac{\wp'(u_t)-\wp'(u_x)}{\wp(u_x)\wp(u_t)}\ u_{xt}+
 \frac{\wp'(u_y)-\wp'(u_t)}{\wp(u_y)\wp(u_t)}\ u_{yt}=0,
$$
which appeared in Sect. 4.4,
possesses four conservation  laws, the first one being
$$
H(u_y, u_t)_x+H(u_t, u_x)_y+H(u_x, u_y)_t=0,
$$
where the function $H(r, s)$ is defined by the equations
$$
H_r=-\zeta^2(s)-\wp(s)-4\wp(s)\frac{\zeta(s)\wp(s)-\zeta(r)\wp(r)}{\wp'(s)-\wp'(r)}, ~~~
H_s=\zeta^2(r)+\wp(r)+4\wp(r)\frac{\zeta(s)\wp(s)-\zeta(r)\wp(r)}{\wp'(s)-\wp'(r)},
$$
here $\zeta$ is the Weierstrass zeta-function: $\zeta'=-\wp$ (we point out that the equations for $H$ are automatically consistent: $H_{rs}\equiv H_{sr}$). Three extra conservation laws are of the form
$$
F(u_x, u_y)_t-F(u_x, u_t)_y=0, ~~~ F(u_y, u_x)_t-F(u_y, u_t)_x=0, ~~~ F(u_t, u_x)_y-F(u_t, u_y)_x=0,
$$
where the function $F(r, s)$ is defined by the equations
$$
F_r=\frac{\wp(r)\wp(s)}{\wp'(r)-\wp'(s)}, ~~~ F_s=\frac{\wp^2(r)}{\wp'(s)-\wp'(r)}-\frac{1}{2}\zeta(r).
$$
Explicitly, one has (see \cite{Moro}, where the function $F$ appeared in a different context):
$$
F(r, s)=\frac{1}{6}\left(\ln \sigma (r-s)+\epsilon \ln \sigma (\epsilon r-s)+\epsilon^2\ln \sigma (\epsilon^2r-s)\right);
$$
here $\epsilon^3=1$ and $\sigma$ is the Weierstrass sigma-function: $\sigma'/\sigma=\zeta$. Notice that  $F$ is real-valued.

\medskip

\noindent The main result of this section is the following

\begin{theorem}
Any integrable quasilinear PDE of the form \eqref{u} possesses four first order conservation
laws.
\end{theorem}

\noindent {\bf Remark 1.} We would like to stress that the converse statement is not true: the existence of four conservation laws does not necessarily imply the integrability. For instance, any Euler-Lagrange equation of the form $(g_{u_x})_x+ (g_{u_y})_y+(g_{u_t})_t=0$, corresponding to the Lagrangian density $ g(u_x, u_y, u_t) $, is manifestly conservative and, moreover,  possesses three extra first order conservation laws
$$
\begin{array}{c}
(u_xg_{u_x}-g)_x+(u_xg_{u_y})_y+(u_xg_{u_t})_t=0, \\
\ \\
(u_yg_{u_x})_x+(u_yg_{u_y}-g)_y+(u_yg_{u_t})_t=0, \\
\ \\
(u_tg_{u_x})_x+(u_tg_{u_y})_y+(u_tg_{u_t}-g)_t=0,
\end{array}
$$
which constitute components of the energy-momentum tensor. On the other hand, as shown in \cite{FKT}, the integrability conditions are very restrictive and reduce to a complicated system of fourth order PDEs for the Lagrangian density $g$, resulting in the finite-dimensionality of the moduli space of integrable Lagrangians.

\medskip

\noindent {\bf Remark 2.} Integrable multi-dimensional equations normally  possess infinite hierarchies of higher order conservation laws, which are generically non-local. The discussion of higher conservation laws  is beyond the scope of our paper.

\medskip

\centerline {\bf Proof of Theorem 2:}

\medskip

\noindent Differentiating Eq. (\ref{conserv-law}) we obtain
\begin{equation*}
    (g_1)_a u_{xx}+ (g_2)_b u_{yy} + (g_3)_c u_{zz}  + ((g_1)_b + (g_2)_a) u_{xy} + ((g_1)_c + (g_3)_a) u_{xz} + ((g_2)_c + (g_3)_b) u_{yz} = 0.
\end{equation*}
Let us introduce the new  variables $s^i, r^i$ via
\begin{align*}
    & r^1 = (g_1)_a, \quad r^2 = (g_2)_b, \quad r^3 = (g_3)_c, \\
    & 2s^1 = (g_2)_c + (g_3)_b, \quad 2s^2 = (g_1)_c + (g_3)_a, \quad 2s^3 = (g_1)_b + (g_2)_a;
\end{align*}
one can verify that these variables automatically satisfy the compatibility conditions
\begin{align}\label{conserv-compat}
     s^1_{aa} &= s^2_{ab} + s^3_{ac} - r^1_{bc}, ~~~~~ 2s^3_{ab} = r^1_{bb} + r^2_{aa}, \notag\\
     s^2_{bb} &= s^1_{ab} + s^3_{bc} - r^2_{ac}, ~~~~~ 2s^1_{bc} = r^2_{cc} + r^3_{bb},\\
     s^3_{cc} &= s^1_{ac} + s^2_{bc} - r^3_{ab}, ~~~~~ 2s^2_{ac} = r^1_{cc} + r^3_{aa}. \notag
\end{align}
For  a relation of the from (\ref{conserv-law}) to be a conservation law of Eq. (\ref{u}) one has to require the existence of a factor $k(a, b, c)$ such that
\begin{align*}
    & r^1 = k f_{11}, ~~ r^2 = k f_{22}, ~~ r^3 = k f_{33}, \\
    & s^1 = k f_{23}, ~~ s^2 = k f_{13}, ~~ s^3 = k f_{12}.
\end{align*}
Substituting this  into the compatibility conditions \eqref{conserv-compat} we arrive at  six linear equations
for the `integrating factor' $k$.
Solving this system  for  the  second order partial derivatives $k_{ij}$ (the variables $a, b, c$ are labelled by indices $1, 2, 3$, respectively), one obtains
\begin{equation}
 k_{ij} = \frac{1}{2}(f_{is}f_{rj}+f_{js}f_{ri}-f_{ij}f_{rs})(kf_{pq, tl}+2k_pf_{tl, q})
    \epsilon^{ptr}\epsilon^{qls}.
\label{kij}
\end{equation}
Here $\epsilon^{ijk}$ is the totally antisymmetric tensor  dual to the volume form
of the metric $f_{ij}$ corresponding to the  equation \eqref{u}, that is,  $\epsilon^{123}=1/\sqrt F,
\epsilon^{213}=-1/\sqrt F,$  etc,  $F={\rm det  }f_{ij}$.  The system (\ref{kij}) appears to be in involution modulo the integrability conditions  (\ref{int-cond}).  Since the variable  $k$ and its first order derivatives $k_a, k_b, k_c$ are not restricted by any additional constraints, there is
 a 4-parameter freedom for the integrating factor. This finishes the proof.

\medskip

 \noindent {\bf Example 5.} Let us write out Eqs. (\ref{kij}) for the dispersionless Hirota equation discussed in Ex. 1. We have $f_{11}=f_{22}=f_{33}=0, \ f_{12}=a_3 c, \ f_{13}=a_2 b, \ f_{23}=a_1 a$, and the substitution into (\ref{kij}) implies
 $$
 k_{ab}=k_{ac}=k_{bc}=0, ~~ k_{aa}=-\frac{3}{a} k_a, \ k_{bb}=-\frac{3}{b} k_b, \ k_{cc}=-\frac{3}{c} k_c.
 $$
This leads to the four values for the integrating factor: $k={\rm const}, \ k=a^{-2}, \ k=b^{-2}, \ k=c^{-2}$, which correspond to the four conservation laws from Ex. 1.

\section{Dispersionless Lax pairs}

In this section we prove that any integrable equation of the form (\ref{u}) possesses a dispersionless Lax pair,
\begin{equation*}
S_t=F(S_x, \, u_x, \, u_y, \, u_t), ~~~
S_y=G(S_x, \, u_x, \, u_y, \, u_t).
\end{equation*}
Let us begin with illustrating examples.

\medskip

\noindent {\bf Example 1.} The dispersionless Hirota equation,
\begin{equation*}
   a_1 u_x u_{yt} + a_2 u_y u_{xt} + a_3 u_t u_{xy} = 0, ~~~ a_1+a_2+a_3=0,
\end{equation*}
possesses the  Lax pair
$$
 \frac{S_t}{u_t}=\mu \frac{S_x}{u_x}, ~~~ \frac{S_y}{u_y}=\lambda\frac{S_x}{u_x};
$$
here the constants $\lambda$ and $\mu$ satisfy a single quadratic constraint $a_1 \lambda \mu+a_2 \mu + a_3 \lambda=0$. Notice that  this  Lax pair is linear in $S$. It was discussed in \cite{Zakharevich} in the context of Veronese webs, and was used to solve the dispersionless Hirota equation via a non-linear Riemann problem.  We will see below that the case $a_1+a_2+a_3\ne 0$ is far more complicated, leading to Lax pairs  parametrized by hypergeometric functions.

\medskip

\noindent {\bf Example 2.} The  dToda singular manifold equation,
$$
u_{xy}-\frac{u_xu_y}{(e^{u_t}-1)(1-e^{-u_t})} u_{tt}=0,
$$
possesses the Lax pair
$$
 {S_t}=F({S_x}/{u_x},  u_t), ~~~ {S_y}=F_2({S_x}/{u_x}, u_t) u_y;
$$
here the function $F(x^1, x^2)$ is defined as
$$
F(x^1, x^2)=-\frac{1}{\lambda}\ln \left(1-\lambda x^1 (1-e^{-x^2})  \right), ~~~ \lambda={\rm const},
$$
$F_2=\partial F / \partial {x^2}$. Although the constant $\lambda$ can be eliminated by a rescaling $S\to \lambda S$,  one can consider the limit as $\lambda \to 0$. This results in a linear Lax pair for the same equation:
$$
 {S_t}=(1-e^{-u_t})\frac{S_x}{u_x}, ~~~ {S_y}=e^{-u_t}\frac{S_x}{u_x} u_y.
$$

\medskip

\noindent {\bf Parametric Lax pairs}. In many cases it turns out to be more convenient to work with parametric Lax pairs,
\begin{equation}
S_x=f(p, u_x, u_y, u_t), ~~ S_y=g(p, u_x, u_y, u_t), ~~ S_t=h(p, u_x, u_y, u_t),
\label{par}
\end{equation}
here $p$ is a parameter. Expressing $p$ from the first equation and substituting into the last two one gets  a Lax pair in the form (\ref{Lax}). Similar parametric Lax pairs appeared previously in the context of the universal Whitham hierarchy \cite{Kr}, see also \cite{O}. The condition for (\ref{par}) to be a Lax pair for Eq.  (\ref{u}) can be derived as follows. Thinking of $p$ as a function of $x, y, t$, and calculating the compatibility conditions $S_{xy}=S_{yx}, \ S_{xt}=S_{tx}, \ S_{yt}=S_{ty}$, one obtains three relations which are linear in $p_x, p_y$ and $p_t$.  It is easy to see that the rank of the matrix at the derivatives $p_x, p_y, p_t$ equals two, so that one can obtain a single relation which does not contain the derivatives of $p$:
\begin{equation}
\begin{array}{c}
(h_ag_p-g_ah_p)u_{xx}+(f_bh_p-h_bf_p)u_{yy}+(g_cf_p-f_cg_p)u_{tt}+\\
\ \\
((f_a-g_b)h_p+h_bg_p-h_af_p)u_{xy}+\\
\ \\
((h_c-f_a)g_p+g_af_p-g_ch_p)u_{xt}+\\
\ \\
((g_b-h_c)f_p+f_ch_p-f_bg_p)u_{yt}=0;
\end{array}
\label{par1}
\end{equation}
this relation must be satisfied identically modulo Eq. (\ref{u}). This requirement leads to the relations
\begin{equation}
\begin{array}{c}
h_ag_p-g_ah_p=kf_{11}, ~~ f_bh_p-h_bf_p=kf_{22}, ~~ g_cf_p-f_cg_p=kf_{33},\\
\ \\
(f_a-g_b)h_p+h_bg_p-h_af_p=2kf_{12}, \\
\ \\
(h_c-f_a)g_p+g_af_p-g_ch_p=2kf_{13},\\
\ \\
 (g_b-h_c)f_p+f_ch_p-f_bg_p=2kf_{23},
\end{array}
\label{6}
\end{equation}
where $k=k(p, a, b, c)$ is the coefficient of proportionality. The set of relations (\ref{6}) can be represented in a compact form
$$
det  \left(
\begin{array}{ccc}
f_p & g_p & h_p\\
df & dg & dh\\
da & db & dc
\end{array}
\right)=kf_{ij}dp^idp^j,
$$
recall that $(u_x, u_y, u_t)=(a, b, c)=(p^1, p^2, p^3)$. We point out  that, by virtue of (\ref{6}), the triple  $(f_p, g_p, h_p)$ satisfies the dispersion relation:
$$
f_{11}f_p^2 + f_{22}g_p^2 + f_{33}h_p^2 + 2f_{12}f_pg_p + 2f_{13}f_ph_p
 + 2f_{23}g_ph_p=0.
$$
Parametric Lax pairs are particularly useful when the equation under study is  symmetric under the interchange of  $x, y, t$:

\medskip

\noindent {\bf Example 3.} Let us consider the  equation
\begin{equation*}
   a_1 u_x u_{yt} + a_2 u_y u_{xt} + a_3 u_t u_{xy} = 0,
\end{equation*}
where, in contrast to Ex. 1,  the constants $a_1, a_2, a_3$ are arbitrary. Without any loss of generality we will normalize them so that $a_1+a_2+a_3=1$. We seek a Lax pair in parametric form
$$
S_x=f(p)u_x, ~~~ S_y=g(p)u_y, ~~~S_t=h(p)u_t.
$$
The corresponding Eq. (\ref{par1}) is
$$
(g-h)f'u_xu_{yt}+(h-f)g'u_yu_{xt}+(f-g)h'u_tu_{xy}=0,
$$
here $'=d/dp$, so that one can set
\begin{equation}
f'=a_1k(f-g)(f-h), ~~~ g'=a_2k(g-f)(g-h), ~~~ h'=a_3k(h-f)(h-g),
\label{fgh1}
\end{equation}
where $k=k(p)$ is a coefficient of proportionality. We point out that the system  (\ref{fgh1}) possesses a conservation law
$$
(g-h)^{a_1}(h-f)^{a_2}(f-g)^{a_3}={\rm const}.
$$
Introducing  the new independent variable $q=(f-g)/(f-h)$ (notice that we have a reparametrization freedom $p\to \varphi (p)$), and using the identity $q'=k(f-g)(g-h)/(f-h)$, one can linearize the system  (\ref{fgh1}),
$$
\frac{df}{dq}=\frac{a_1}{1-q}(f-h), ~~~ \frac{dg}{dq}=a_2(h-f), ~~~ \frac{dh}{dq}=\frac{a_3}{q}(f-h),
$$
Noticing that $f-h=(1-q)^{-a_1}q^{-a_3}$,  one can reduce these equations  to hypergeometric integrals,
$$
\frac{df}{dq}=a_1(1-q)^{-a_1-1}q^{-a_3}, ~~~ \frac{dg}{dq}=-a_2(1-q)^{-a_1}q^{-a_3}, ~~~ \frac{dh}{dq}=a_3(1-q)^{-a_1}q^{-a_3-1}.
$$
Explicitly, one arrives at the `reparametrized' Lax pair
\begin{equation}
\frac{S_x}{u_x}=f(q), ~~~ \frac{S_y}{u_y}=g(q), ~~~ \frac{S_t}{u_t}=h(q),
\label{hyper}
\end{equation}
where
$$
f=F(q), ~~~ g= F(q)-(1-q)^{-a_1}q^{1-a_3}, ~~~ h=F(q)-(1-q)^{-a_1}q^{-a_3},
$$
and $F(q)$ is a the hypergeometric function:
$$
F(q)=\frac{a_1}{1-a_3}q^{1-a_3} \ _{2}F_1(a_1+1, 1-a_3; 2-a_3; q).
$$

\noindent In the symmetric case $a_1=a_2=a_3=\frac{1}{3}$ we obtain a Lagrangian equation
$$
 u_x u_{yt} +  u_y u_{xt} +  u_t u_{xy} = 0,
 $$
which corresponds to the Lagrangian density $u_xu_yu_t$. Its parametric Lax pair was calculated in \cite{FKT} in the form (here the parameter is denoted by $z$),
\begin{equation}
\frac{S_x}{u_x}=\zeta(z), ~~~
\frac{S_y}{u_y}=\zeta(z)+\frac{\wp'(z)+\lambda}{2\wp(z)}, ~~~
 \frac{S_t}{u_t}=\zeta(z)+\frac{\wp'(z)-\lambda}{2\wp(z)},
 \label{z}
 \end{equation}
where $\wp (z)$ is the Weierstrass $\wp$-function,
$(\wp')^2=4\wp^3+\lambda^2$ (notice that $g_2=0, \ g_3=-\lambda^2$), and
$\zeta (z)$ is the corresponding zeta-function: $\zeta' =-\wp $. The Lax pair (\ref{z}) transforms into (\ref{hyper}) via a substitution
$q=(\wp'(z)+\lambda)/(\wp'(z)-\lambda)$.

\medskip

\noindent {\bf Example 4.} The  equation
$$
u_t(u_x^3-u_y^3)u_{xy}+ u_y(u_t^3-u_x^3)u_{xt}+ u_x(u_y^3-u_t^3)u_{yt}=0
$$
possesses the parametric Lax pair
$$
S_x=G\left(\frac{u_x}{p}\right), ~~ S_y=G\left(\frac{u_y}{p}\right), ~~ S_t=G\left(\frac{u_t}{p}\right),
$$
where the function $G(s)$ satisfies  the equations $G'=\frac{1}{s^3-1}$.
Explicitly, one has
$$
G(s)=\frac{1}{3}\left(\ln (s-1)+\epsilon \ln (s-\epsilon)+\epsilon^2 \ln (s-\epsilon^2)\right).
$$

\medskip

\noindent {\bf Example 5.} The  equation
$$
\frac{\wp'(u_x)-\wp'(u_y)}{\wp(u_x)\wp(u_y)}\ u_{xy}+
\frac{\wp'(u_t)-\wp'(u_x)}{\wp(u_x)\wp(u_t)}\ u_{xt}+
 \frac{\wp'(u_y)-\wp'(u_t)}{\wp(u_y)\wp(u_t)}\ u_{yt}=0
$$
possesses the parametric Lax pair
$$
S_x=F(p, u_x), ~~ S_y=F(p, u_y), ~~ S_t=F(p, u_t),
$$
where the function $F(r, s)$ is defined by the equations
$$
F_r=\frac{\wp(r)\wp(s)}{\wp'(r)-\wp'(s)}, ~~~ F_s=\frac{\wp^2(r)}{\wp'(s)-\wp'(r)}-\frac{1}{2}\zeta(r).
$$
Explicitly, one has
$$
F(r, s)=\frac{1}{6}\left(\ln \sigma (r-s)+\epsilon \ln \sigma (\epsilon r-s)+\epsilon^2\ln \sigma (\epsilon^2r-s)\right),
$$
compare with Ex. 4 of Sect. 5. Similar parametric Lax pairs can be constructed for all integrable examples obtained in Sect. 4.4.

\medskip

\noindent The main result of this Section is the following

\begin{theorem}
Any integrable equation  of the form \eqref{u}  possesses a dispersionless
Lax pair. Furthermore, the existence of a dispersionless Lax pair is equivalent to the existence of an infinity of hydrodynamic reductions and, thus, is necessary and sufficient for the integrability.
\end{theorem}

\medskip

\centerline{\bf Proof:}

\medskip

\noindent Setting $S_x=\xi, \ u_x=a,\ u_y=b, \ u_t=c$, and calculating the consistency condition for the Lax pair (\ref{Lax}), one obtains
$$
\begin{array}{c}
 S_{ty}-S_{yt}= (F_{\xi}G_a-G_{\xi}F_a)u_{xx}+F_bu_{yy}-G_cu_{tt} +\\
 \ \\
 (F_a+F_{\xi}G_b-G_{\xi}F_b)u_{xy}-(G_a+G_{\xi}F_c-F_{\xi}G_c)u_{xt}+(F_c-G_b)u_{yt}.
 \end{array}
 $$
 Since this expression has to vanish modulo Eq.  (\ref{u}), one arrives at the relations
 $$
\begin{array}{c}
F_{\xi}G_a-G_{\xi}F_a=kf_{11}, ~~ F_b=kf_{22}, ~~ G_c=-kf_{33},\\
 \ \\
 F_a+F_{\xi}G_b-G_{\xi}F_b=2kf_{12}, ~~ G_a+G_{\xi}F_c-F_{\xi}G_c=-2kf_{13}, ~~ F_c-G_b=2kf_{23},
 \end{array}
 $$
where $k=k(\xi, a, b, c)$ is the coefficient of proportionality. The last five relations imply the expressions for the derivatives of $F$ and $G$  in the form
\begin{align}\label{FG}
     F_a &= 2kf_{12}+G_{\xi}kf_{22}+F_{\xi}(kf_{23}-p), ~~~ &G_a& =-2kf_{13}-F_{\xi}kf_{33}-G_{\xi}(kf_{23}+p) , \notag\\
     F_b &= kf_{22}, ~~~ &G_b& = -kf_{23}+p,\\
     F_c &= kf_{23}+p, ~~~ &G_c &= -kf_{33}, \notag
\end{align}
where $p=p(\xi, a, b, c)$ is yet another auxiliary function. Substituting these expressions  into the first relation, one can see that $F_{\xi}$ and $G_{\xi}$ have to satisfy the dispersion relation,
\begin{equation}
f_{11} + f_{22}G_{\xi}^2 + f_{33}F_{\xi}^2 + 2f_{12}G_{\xi} + 2f_{13}F_{\xi}
 + 2f_{23}F_{\xi}G_{\xi}=0.
\label{disp1}
\end{equation}
To close the system (\ref{FG}) -- (\ref{disp1}) one proceeds as follows. Calculating the consistency conditions for Eqs. (\ref{FG}), $F_{ab}=F_{ba}$,  $G_{ab}=G_{ba}$,  etc, six conditions altogether, and differentiating the dispersion relation (\ref{disp1}) by $a, b, c$ and $\xi$, one obtains ten relations which can be solved for $F_{\xi \xi}, G_{\xi \xi}$ and the first order derivatives of $k$ and $p$. The resulting system is in involution if and only if the integrability conditions of Sect. 2 are satisfied. This finishes the proof of Theorem 3.

\section{Differential-geometric aspects of the integrability conditions}

As explained in Sect. 3,  the  differential-geometric picture behind equations of the form (\ref{2nd}) is the projective space $P^n$ with coordinates $p^1, ..., p^n$ supplied with the conformal structure
 $f_{ij}dp^idp^j$. The equivalence group $SL(n+1, R)$ acts by projective transformations of $P^n$. Two equations are equivalent if and only if  conformal classes of the corresponding metrics are projectively equivalent.

Let us consider Lagrangian equations of the form (\ref{2nd}) which arise as the Euler-Lagrange equations from  the functionals $\int g({\bf p}) d{\bf x}$
 where the density $g(p^1, ..., p^n)$  depends on the first order derivatives $p^i=u_{x_i}$ only.  In this case the coefficient matrix $f_{ij}$ is the Hessian matrix of  $g$. Our first remark, which is true in any dimension, is that the class of Lagrangian systems is invariant under the action of the equivalence group defined by Eqs. (\ref{tp}), (\ref{tA}). One can show that the extension of the projective action (\ref{tp}) to the Lagrangian density $g$ is given by the formula
 \begin{equation}
 \tilde g=\frac{g}{1+{\bf p}C^{-1}b}.
 \label{tf}
 \end{equation}
The projective  invariance of the class of Lagrangian systems can be seen as follows.
 Let us consider the Lagrangian density $g({\bf p})$ as the equation  of a hypersurface in $P^{n+1}$ defined as  $p^{n+1}=g({\bf p})$. The second fundamental form of this hypersurface coincides with the conformal class of the second differential $d^2g$.  The transformation (\ref{tp}), (\ref{tf}) is a projective transformation in $P^{n+1}$. Thus, the fact that $d^2g$ transforms into $d^2\tilde g$ (up to a conformal factor) is nothing but the well-known projective invariance of the second fundamental form.

 \medskip

A geometric characterization of linearizable equations (\ref{2nd}) is provided in Sect. 7.1 (Theorem 4). To be precise, we will be interested in  those equations  which can be linearized by a transformation from the equivalence group.

\medskip

A simple tensorial characterization of linearizable and Lagrangian equations is given in Sect. 7.2. The answer is formulated in terms of the tensor $a_{ijk}$ and the projectively flat connection $\nabla$ with Christoffel's symbols $\Gamma^i_{jk}=s_j\delta^i_k+s_k\delta^i_j$ which are naturally assocated with the conformal structure $f_{ij}dp^idp^j$.

\medskip

Invariant differential-geometric formulation of the integrability conditions (\ref{int-cond}) derived in Sect. 2 is provided in Sect. 7.3. This involves the tensor $a_{ijk}$ and its covariant derivative
with respect to  $\nabla$.

\medskip

Finally,  a simple differential-geometric characterization of conformal structures corresponding to integrable equations is proposed in Sect. 7.4.

\subsection{Linearizable equations and quadratic line complexes}

Before formulating the main result, let us summarize  the properties of linear (linearizable)  equations.

 \noindent {\bf Example 1.} Consider the $3$-dimensional linear wave equation,
 $$
 u_{tt}=u_{xx}+u_{yy};
 $$
 notice first  that it is Lagrangian with the quadratic Lagrangian density $g=u_x^2+u_y^2-u_t^2$. The associated conformal structure in $P^3$ corresponds to the standard Lorentzian metric $(dp^1)^2+(dp^2)^2-(dp^3)^2$; here $p^1=u_x, \ p^2=u_y, \ p^3=u_t$. This conformal structure possesses a $3$-parameter family of null lines
 defined by the equations
 \begin{equation}
 p^1=\alpha p^3+\beta, ~~~ p^2=\gamma p^3+\delta,
 \label{line}
 \end{equation}
 where the constants
 $\alpha, \beta, \gamma, \delta$ satisfy a single quadratic constraint $\alpha^2+\gamma^2=1$. The last property  can be reformulated in a projectively  invariant way as follows. Recall that a $3$-parameter family of lines in $P^3$ is called a {\it line complex}. A complex is said to be {\it quadratic} if it is defined by a single quadratic relation among the Plucker coordinates in the space of lines (which is identified with the  Plucker quadric in $P^5$).  In the parametrization (\ref{line}), the Plucker coordinates are
  $$
 (1 : \alpha : \beta : \gamma :  \delta : \alpha \delta-\gamma \beta).
 $$
Fixing a point  in $P^3$ with coordinates $p^1_0, \ p^2_0,\  p^3_0$, the  lines of the complex passing  through this point generate a quadratic cone with the equation
$(p^1-p^1_0)^2+(p^2-p^2_0)^2=(p^3-p^3_0)^2$; these cones are nothing but the null cones of the corresponding conformal structure. Introducing in $P^3$ homogeneous coordinates $q^0:q^1:q^2:q^3$ via $p^i=q^i/q^0$,
one can see that the intersection of  null cones with the plane at infinity (defined by the equation
$q^0=0$)  is the conic $(q^1)^2+(q^2)^2=(q^3)^2$. Thus, all quadratic cones of our complex pass through one and the same plane conic  (complexes of this type have the Segre symbol $[(222)]$, see \cite{Jessop, Avritzer}).
Summarizing, we see that

\noindent (a) the linear wave equation is Lagrangian;

\noindent  (b) the corresponding conformal structure possesses a three-parameter family of null lines which form a quadratic complex with  the Segre symbol $[(222)]$ (equivalently, quadratic cones of the complex pass through one and the same plane conic).

\noindent
 Reformulated in these terms, both properties are manifestly projectively invariant, and hold for  arbitrary equations related to the linear wave equation via the action of the  equivalence group.

\noindent {\bf Example 2.} Let us consider the  equation
$$
(1-u_x^2-u_y^2)u_{tt}-u_t^2(u_{xx}+u_{yy})+2u_t(u_xu_{xt}+u_yu_{yt})=0,
$$
which is Lagrangian with the Lagrangian density $g=\frac{u_x^2+u_y^2-1}{u_t}$. The corresponding conformal structure,
$$
(1-(p^1)^2-(p^2)^2)(dp^3)^3-(p^3)^2((dp^1)^2+(dp^2)^2)+2p^3(p^1dp^1dp^3+p^2dp^2dp^3),
$$
possesses a $3$-parameter family of null lines (\ref{line}) specified by a single  quadratic relation
$\beta^2+\delta^2=1$. This defines a quadratic complex whose null cones pass through one and the same planar conic defined by the equation $(q^1)^2+(q^2)^2=(q^0)^2$ in the plane $q^3=0$ (recall that $q^i$ are homogeneous coordinates in $P^3$). The equation  linearizes (to the wave equation from  Ex. 1)  under the transformation
$$
\tilde u=-t, ~~~ \tilde t =-u, ~~~ \tilde y= y, ~~~ \tilde x=x,
$$
which generates a projective transformation of the derivatives,
$$
\tilde u_{\tilde t}=\frac{1}{u_t}, ~~~ \tilde u_{\tilde y}=\frac{u_y}{u_t}, ~~~ \tilde u_{\tilde x}=\frac{u_x}{u_t}.
$$
This extends to the transformation of  the Lagrangian densities as  $\tilde g={g}/{u_t}$. One can verify  that the quadratic Lagrangian density $g$ of the linear wave equation  transforms to the density $\tilde g$ of the linearizable equation from  Ex. 2. Geometrically, this transformation is nothing but a projective transformation which sends the plane $q^3=0$ to the plane at infinity.

Our observations are summarized in the following theorem which, in fact, holds in any dimension.

 \begin{theorem}
 The following conditions are equivalent:

 \noindent (1) Eq. (\ref{2nd}) is linearizable by a transformation from the equivalence group.

 \noindent (2) Eq.  (\ref{2nd}) is Lagrangian with the Lagrangian density
 $g=\frac{Q({\bf p})}{l({\bf p})}$
 where $Q$ and $l$ are arbitrary quadratic and linear forms in $p^1, ..., p^n$, respectively (not necessarily homogeneous).

 \noindent (3) The conformal structure  $f_{ij}dp^idp^j$ possesses a complex (that is, a $(2n-3)$-parameter family) of null lines whose quadratic cones pass through a stationary hyperplane quadric.
 For $n=3$ these conditions are equiavalent to the requirement that the complex has Segre symbol [(222)].

\end{theorem}

\centerline {\bf Proof:}

The equivalence of (1) and (2) can be seen as follows. Suppose that Eq. (\ref{2nd}) is linearizable. Then the corresponding conformal structure $f_{ij}dp^idp^j$ is transformable to a constant coefficient form. Since any constant coefficient linear system is Lagrangian with a quadratic Lagrangian density $g$, and the class of Lagrangian systems is projectively invariant, any linearizable equation is necessarily Lagrangian. Applying the projective transformation (\ref{tp}), (\ref{tf}) to a quadratic Lagrangian density, one obtains a density of the form  $g=\frac{Q({\bf p})}{l({\bf p})}$
where $Q$ and $l$ are quadratic and linear expressions in $p^1, ..., p^n$, respectively. Conversely, given a Lagrangian density of the form $g=\frac{Q({\bf p})}{l({\bf p})}$, and applying any projective transformation which has the linear form $l({\bf p})$ in the denominator, one obtains a purely quadratic Lagrangian density which gives rise to a linear equation. This establishes the equivalence of (1) and (2).

The implication (1) $\implies$ (3) is straightforward: any constant coefficient conformal structure possesses a complex of null lines whose quadratic cones pass through a stationary quadric belonging to the hyperplane at infinity. Conversely, consider an equation whose conformal structure possesses a quadratic complex of null lines such that all null cones  pass through a stationary quadric belonging to a stationary hyperplane $H$. Applying a projective transformation which sends $H$ to a hyperplane at infinity, we obtain a linear equation with constant coefficients. In the case $n=3$ one can also refer to the Proposition 4.3. of \cite{Avritzer} which implies that, up to projective equivalence, there exists a unique quadratic complex with the Serge symbol [(222)].

\medskip
The constraints on the complex are crucial for the linearizability.

\noindent {\bf  Example 3.}  Let us consider the dispersionless Hirota equation
$$
a_1u_xu_{yt}+a_2u_yu_{xt}+a_3u_tu_{xy}=0, ~~~~~~ a_1+a_2+a_3=0;
$$
we point out that the equation  is integrable for any values of  constants, not necessarily satisfying this relation. The corresponding conformal structure
$$
a_1p^1dp^2dp^3+a_2p^2dp^1dp^3+a_3p^3dp^1dp^2
$$
possesses a three-parameter family of null lines which form a quadratic complex defined by the equation
$a_1\beta \gamma+a_2\alpha \delta =0$. This complex is not of the Segre type $[(222)]$, therefore, the equation is not linearizable. In fact, it is easy to see that it is not Lagrangian.

\noindent {\bf  Example 4.} Let us consider the equation
$$
u_{xx}+(u_y^2-1)u_{tt}-2u_yu_tu_{yt}+u_t^2u_{yy}=0;
$$
one can show that it  is not integrable, and not Lagrangian. The corresponding conformal structure
$$
(dp^1)^2+((p^2)^2-1)(dp^3)^2-2p^2p^3dp^2dp^3+(p^3)^2(dp^2)^2
$$
possesses a three-parameter family of null lines which form a quadratic complex defined by the equation $\alpha^2+\delta^2=1$, which is again not of the Segre type $[(222)]$.

\noindent {\bf Remark.} The condition for a conformal structure $f_{ij}dp^idp^j$ in $P^n$ to possess a complex
of null lines is equivalent to a simple differential-geometric constraint
\begin{equation}
\partial_{(k}f_{ij)}=\varphi_{(k}f_{ij)};
\label{sym}
\end{equation}
here $\partial_k=\partial_{p^k}$,  $\varphi_k$ is a covector, and  brackets denote a complete symmetrization in $i, j, k$. Contracting (\ref{sym}) with $f^{ij}$ we obtain
$$
\varphi_k=
\frac{1}{n+2} f^{pq}(\partial_{k}f_{pq}+2\partial_pf_{qk}).
$$
The condition (\ref{sym}) characterizes  conformal structures coming from quadratic line complexes, see e.g. \cite{Akivis,  Safaryan} and references therein.   Thus, the identity (\ref{sym}) is necessary (although not sufficient) for the linearizability. An invariant characterization of linearizable equations is provided in Sect. 7.2. below.

\bigskip

\subsection{Tensorial characterization of linearizable and Lagrangian equations}

The results of this section are valid in any dimension, and provide a simple differential-geometric criterion of the linearizability for equations of the form (\ref{2nd}).

\medskip

\noindent {\bf Lemma}. {\it An equation of the form (\ref{2nd}) is linearizable by a transformation from the equivalence group $SL(n+1, R)$ iff there exists a flat connection $\nabla$ with Christoffel symbols $\Gamma^i_{jk}=s_j\delta^i_k+s_k\delta^i_j$ such that
\begin{equation}
\nabla _kf_{ij}=c_kf_{ij};
\label{nabla}
\end{equation}
(connections satisfying Eq. (\ref{nabla}) are known as Weyl connections).}

\medskip

\centerline{\bf Proof:}

\medskip

\noindent The necessity is straightforward: given a linear equation with constant coefficients $f_{ij}$, we immediately arrive at (\ref{nabla}) where $\nabla$ is  a flat connection with  zero Christoffel symbols, $\nabla_k=\partial_k=\partial/\partial_{p^k}$,  and
$c_k=0$. Applying a transformation from the equivalence group (which acts  projectively  on the space $P^n$ with coordinates $p^1, ..., p^n$) to a flat connection $\nabla$, we will obtain a flat connection with nonzero Christoffel symbols of the form $\Gamma^i_{jk}=s_j\delta^i_k+s_k\delta^i_j$, which still satisfies  Eqs. (\ref{nabla}). Moreover, the condition (\ref{nabla}) is manifestly invariant under  rescalings $f_{ij}\to \tau f_{ij}$ (under such rescalings, the covector $c_k$ transforms to $c_k+\nabla_k \ln \tau$).

Conversely, suppose a connection $\nabla$ has Christoffel symbols of the form $\Gamma^i_{jk}=s_j\delta^i_k+s_k\delta^i_j$ (connections of this form are known as projectively flat: their geodesics are straight lines). If, in addition, $\nabla$ is flat (has zero curvature tensor), there exists a {\it projective} transformation bringing Christoffel symbols to zero. In the new coordinates Eq. (\ref{nabla}) will take the form $\partial _{k}f_{ij}=c_kf_{ij}$, which implies that the coefficient matrix $f_{ij}$  is proportional to a constant matrix. This finishes the proof.

\medskip

Relations (\ref{nabla}) lead to explicit tensorial constraints for $f_{ij}$ as follows. Taking into account that
$\Gamma^i_{jk}=s_j\delta^i_k+s_k\delta^i_j$ one can rewrite (\ref{nabla}) as
\begin{equation}
\partial_{k}f_{ij}=(c_k+2s_k)f_{ij}+s_if_{kj}+s_jf_{ki}.
\label{nabla1}
\end{equation}
Contacting Eqs. (\ref{nabla1}) with the inverse matrix $f^{ij}$ one arrives at  the relations
$$
f^{ij}\partial_kf_{ij}=nc_k+2(n+1)s_k, ~~~ f^{ij}\partial_jf_{ik}=c_k+(n+3)s_k,
$$
with a double summation over $i$ and $j$. This implies
\begin{equation}
\begin{array}{c}
s_k=\frac{f^{ij}}{(n+2)(1-n)}\left(\partial_kf_{ij}-n\partial _jf_{ik}\right), \\
\ \\
c_k=\frac{f^{ij}}{(n+2)(n-1)}\left((n+3)\partial_kf_{ij}-2(n+1)\partial _jf_{ik}\right).
\end{array}
\label{cs}
\end{equation}
Given an arbitrary conformal structure $f_{ij}dp^idp^j$ in $P^n$, let us introduce the tensor
$$
a_{ijk}=\partial_kf_{ij}-(c_k+2s_k)f_{ij}-s_if_{kj}-s_jf_{ki},
$$
here $s_k, c_k$ are the same as in Eqs. (\ref{cs}). One can readily verify the apolarity relations
$f^{ij}a_{ijk}=0, \ f^{ij}a_{ikj}=0$. Thus, we can formulate the following

\medskip

\noindent {\bf Proposition 1.} { \it Equation (\ref{2nd}) is linearizable by a transformation from the equivalence group $SL(n+1, R)$ iff the corresponding conformal structure satisfies the following two properties:

\noindent (1) the tensor $a_{ijk}$ vanishes;

\noindent (2) the connection $\Gamma^i_{jk}=s_j\delta^i_k+s_k\delta^i_j$ is flat; a simple calculation shows that this condition is equivalent to  $\partial_js_i-s_is_j=0$.}

\medskip

\noindent {\bf Remark 1.} In a somewhat different form, the tensor $a_{ijk}$ appeared previously in \cite{Safaryan} in the study of manifolds of quadratic cones in $P^n$. It was proved that the vanishing of $a_{ijk}$ alone implies the existence of a stationary hyperquadric $Q^{n-1}\subset P^n$ such that all cones of the family are tangential to $Q^{n-1}$. Clearly, all such conformal structures are projectively equivalent, and can be brought to a canonical form
$$
[({\bf p}, {\bf p})-1](d{\bf p}, d{\bf p})-({\bf p}, d{\bf p})^2=0,
$$
where ${\bf p}=(p^1, ..., p^n)$ are coordinates in $P^n$, and $(\ , \ )$ is the standard scalar product. The null cones of this conformal structure are tangential to a unit sphere centered at the origin. The corresponding second order equation takes the form
$$
[(\nabla u)^2-1]\triangle u-(\nabla u) H (\nabla u)^t=0,
$$
where $\nabla u=(u_{x^1}, ..., u_{x^n})$ is the gradient of $u$,  $\triangle$ is the Laplacian, and $H$ is the Hessian matrix of $u$. In the three-dimensional case we arrive at the equation
$$
\begin{array}{c}
(u_y^2+u_t^2-1)u_{xx}+(u_x^2+u_t^2-1)u_{yy}+(u_x^2+u_y^2-1)u_{tt}\\
\ \\
-2(u_xu_yu_{xy}+u_xu_tu_{xt}+u_yu_tu_{yt})=0;
\end{array}
$$
notice that it is Lagrangian: the corresponding  Lagrangian density $\sqrt {1- u_x^2-u_y^2-u_t^2}$  governs minimal hypersurfaces $z=u(x, y, t)$ in the Lorentzian space with the metric
$dx^2+dy^2+dt^2-dz^2$. We have verified that this equation {\it does not}  satisfy the integrability conditions of Sect. 2.

\medskip

\noindent {\bf Remark 2.} Another result of \cite{Safaryan} (also formulated in different terms) states that, imposed simultaneously,  conditions (1) and (2) imply the existence of a stationary hyperplane $H$ in $P^n$ with a stationary quadric $Q^{n-2}\subset H$ such that all cones of the  family pass through $Q^{n-2}$. This provides an alternative  projectively-invariant characterization of conformal sructures corresponding to linearizable equations:  the linearizing transformation is any projective transformation which sends $H$ to the hyperplane at infinity (see Sect. 7.2).

\medskip

The tensor $a_{ijk}$ provides a simple characterization of Lagrangian equations:

\medskip

\noindent {\bf Proposition 2.} { \it Equation  (\ref{2nd}) is Lagrangian iff the corresponding conformal structure satisfies the following two properties:

\noindent (1) the tensor $a_{ijk}$ is totally symmetric; in fact, since $a_{ijk}$ is manifestly symmetric in the first two indices, it is sufficient to require $a_{ijk}=a_{ikj}$,

\noindent (2) the covector $s_i$ is a gradient: $\partial_js_i=\partial_is_j$;}

\noindent  these conditions are obtained by weakening the corresponding conditions of  Proposition 1 (recall that any linearizable equation is automatically Lagrangian).

\medskip

\centerline{\bf Proof:}

\medskip

\noindent To show that a given equation is  Lagrangian, one has to find an integrating factor $\tau$ such that
the matrix $\tau f_{ij}$ is the Hessian matrix of a function, equivalently,  $\partial_k (\tau f_{ij})=\partial_j (\tau f_{ik})$, which gives
$$
\frac{\partial_k\tau}{\tau} f_{ij}+\partial_kf_{ij}=\frac{\partial_j\tau}{\tau} f_{ik}+\partial_jf_{ik}.
$$
Contracting this expression with $f^{ij}$ one gets
\begin{equation}
\frac{\partial_k\tau}{\tau}=\frac{f^{ij}}{n-1}\left(\partial_jf_{ik}-\partial_kf_{ij}\right).
\label{pgrad}
\end{equation}
Substituting this back into the previous equation one obtains the relation
$$
\partial_kf_{ij}+f_{ij}\frac{f^{pq}}{n-1}\left(\partial_qf_{pk}-\partial_kf_{pq}\right)=
\partial_jf_{ik}+f_{ik}\frac{f^{pq}}{n-1}\left(\partial_qf_{pj}-\partial_jf_{pq}\right),
$$
which is identical with $a_{ijk}=a_{ikj}$. Finally, the right hand side of (\ref{pgrad}) must be a gradient. Since the expression $f^{ij}\partial_kf_{ij}$ is automatically a gradient by virtue of the identity
$$
{\rm tr} F^{-1}\partial_kF=\partial_k\ln \det F,
$$
one has to require that $f^{ij}\partial_jf_{ik}$ is a gradient. This, however, is equivalent to the requirement that the covector $s_i$ must be a gradient.  This finishes the proof.

\medskip

Finally, we have the following

\medskip

\noindent {\bf Proposition 3.} { \it  Conformal structure $f_{ij}dp^idp^j$ in $P^n$  possesses a quadratic complex of null lines iff  the symmetrized  tensor  $a_{ijk}$ vanishes:}
$$
a_{(ijk)}=0.
$$
Indeed,  the condition $a_{(ijk)}=0$ is identical to (\ref{sym}).

\bigskip

\subsection{Tensorial formulation of the integrability conditions}

Let us begin with a general differential-geometric digression. Given a metric $f_{ij}$ and a connection $\hat \nabla$ with Christoffel symbols $\hat \Gamma^i_{jk}$ on a $n$-dimensional manifold, let us introduce the following objects:

\noindent --- covectors $s_k$ and $c_k$:
$$
\begin{array}{c}
s_k=\frac{f^{ij}}{(n+2)(1-n)}\left(\hat \nabla_kf_{ij}-n\hat \nabla _jf_{ik}\right), \\
\ \\
c_k=\frac{f^{ij}}{(n+2)(n-1)}\left((n+3)\hat \nabla_kf_{ij}-2(n+1)\hat \nabla _jf_{ik}\right);
\end{array}
$$

\noindent --- tensor $a_{ijk}$:
$$
a_{ijk}=\hat \nabla_kf_{ij}-(c_k+2s_k)f_{ij}-s_if_{kj}-s_jf_{ki};
$$

\noindent --- tensor $a_{ijkl}$:
$$
a_{ijkl}=\hat \nabla_la_{ijk}-(c_l+3s_l)a_{ijk}-s_ia_{ljk}-s_ja_{ilk}-s_ka_{ijl}.
$$
The importance of these objects is explained by their transformation properties: suppose that the metric $f_{ij}$ and the connection $\hat \nabla$ are allowed to vary within their  conformal and projective classes, respectively, that is,
$$
 f_{ij}\to \varphi f_{ij}, ~~~  \hat \Gamma^i_{jk}\to \hat \Gamma^i_{jk}+\psi_k\delta^i_j+\psi_j\delta^i_k.
$$
One can readily verify the transformation properties
$$
 s_k \to s_k-\psi_k, ~~~  c_k\to c_k+\frac{\partial_k\varphi}{\varphi}, ~~~
 a_{ijk}\to \varphi a_{ijk}, ~~~  a_{ijkl}\to \varphi a_{ijkl},
$$
which, in particular, imply that the new Christoffel symbols,
$$
\Gamma^i_{jk}=\hat \Gamma^i_{jk}+s_k\delta^i_j+s_j\delta^i_k,
$$
give rise to a well-defined affine connection $\nabla$ which depends neither on the choice of $\hat \Gamma^i_{jk}$ in its projective class, nor on the conformal factor $\varphi$. The expressions for $a_{ijk}$ nd $a_{ijkl}$ compactify to
$$
a_{ijk}=\nabla_kf_{ij}-c_kf_{ij}, \qquad a_{ijkl}= \nabla_la_{ijk}-c_la_{ijk}.
$$
In our context, $n=3$, $f_{ij}$ is a conformal structure in $P^3$,  and the  projective structure  is associated with the projective class of a  flat connection:  $\hat \Gamma^i_{jk}=0$ (indeed, only projective transformations preserve the projective class of a flat connection). Thus, $\hat \nabla_k=\partial/\partial p^k$, and the connection $\nabla$ is given by
$$
\Gamma^i_{jk}=s_k\delta^i_j+s_j\delta^i_k;
$$
notice that, although $\nabla$ is manifestly projectively flat, is does not need to be flat (that is, have zero curvatute tensor) in general. The tensors $a_{ijk}$,  $a_{ijkl}$ and the affine connection $\nabla$ constitute a complete set of projective  invariants of the conformal structure $f_{ij} dp^idp^j$.
For $n=3$ the integrability conditions (\ref{int-cond}) can be formulated as follows (we begin with the Lagrangian case, which is computationally simpler):

\medskip

\noindent {\bf Integrability conditions in the Lagrangian case}

\medskip

\noindent  In the Lagrangian case the tensor $a_{ijk}$ is totally symmetric, and the integrability conditions take the form
\begin{equation}
 \begin{aligned}
   \partial_j s_{i} - s_i s_j =
        -\frac{1}{20} a_{tm\mu}a_{rs\nu}f_{iq}f_{jp}f^{\mu
\nu}\varepsilon^{ptr}\varepsilon^{qms}
-\frac{3}{20} a_{ipr}a_{jqs}f^{pq}f^{rs},
\end{aligned}
\label{Lint1}
\end{equation}
and
\begin{equation}
 \begin{aligned}
    a_{ijkl} = &\frac{9}{10}
                Sym~
            a_{ijp}a_{klq}f^{pq}\\
        &-Sym \left( \frac{9}{20}f_{sl}
            a_{ipq} a_{jtm} f_{rk} +
            \frac{3}{2}a_{ltm} a_{kpq} f_{ri} f_{sj}
          -\frac{3}{20}a_{tm\mu} a_{rs\nu}
              f_{iq} f_{jp} f_{kl} f^{\mu\nu}\right) \varepsilon^{ptr} \varepsilon^{qms},
\end{aligned}
\label{Lint2}
\end{equation}
respectively. Here $Sym$ denotes a complete symmetrization in $i, j, k$,
$$
Sym~T_{ijk}=\frac{1}{3!}\sum_{\sigma \in S_3}T_{\sigma{(i)} \sigma{(j)} \sigma{(k)}},
$$
and $\epsilon^{ijk}$ is the totally antisymmetric tensor  dual to the volume form
of the metric $f_{ij}$, that is,  $\epsilon^{123}=1/\sqrt F,
\epsilon^{213}=-1/\sqrt F,$  etc,  $F={\rm det  }f_{ij}$. This provides yet another form of the integrability conditions in the Lagrangian case, compare with \cite{FKT}.  Recall that linearizable equations are characterized by the relations $a_{ijk}=0$,
$\partial_j s_i - s_i s_j = 0$ (Prop. 2 of Sect. 7.2), which clearly annihilate both of the conditions (\ref{Lint1}), (\ref{Lint2}). This is in agreement with the obvious fact that any linearizable equation is automatically integrable.

\medskip

\noindent {\bf Integrability conditions in the general  case}

\medskip

\noindent In the general case the tensor $a_{ijk}$ is no longer symmetric (only in the first two indices), and the integrability conditions become considerably more complicated. Thus, the analogue of Eq. (\ref{Lint1}) takes the form
\begin{equation}
\begin{aligned}
    \partial_js_{i} - s_i s_j =
        -&\frac{1}{20}(
            2 a_{{\mu}tm} a_{{\nu}rs} +
            2a_{{\mu}mt} a_{{\nu}sr}  -
            3 a_{tm{\mu}} a_{rs{\nu}}
           )f_{iq} f_{jp} f^{{\mu}{\nu}} \varepsilon^{ptr} \varepsilon^{qms}\\
         -&\frac{1}{20}(
            6 a_{ipr} a_{jqs} -
            5 a_{pri} a_{qsj} +
            a_{ipr} a_{qsj} +
            a_{pri} a_{jqs}
           )f^{pq} f^{rs};
\end{aligned}
\label{int1}
\end{equation}
one can show that the right hand side of (\ref{int1}) is symmetric with respect to $i$ and $j$, so that
$\partial_js_{i}=\partial_is_{j}$.  This means that, for integrable equations, the covector $s_i$ must be a gradient.  In this case the left hand side of Eq. (\ref{int1}) can be represented in the form
$\partial_js_{i} - s_i s_j = \frac{1}{n-1}R_{ij}$ where $R_{ij}$ is the Ricci tensor on $\nabla$.
The analogue of Eq. (\ref{Lint2}) takes the form
\begin{align}\label{int2}
a_{ijkl} = - \frac{1}{20}{Sym}(
        &( 8 a_{klq} + 44  a_{qkl} - 70  a_{qlk} ) a_{ijp} +
        (64 a_{klq} - 82  a_{qkl} + 30  a_{qlk} ) a_{pij} - 12 a_{pkj} a_{qil} ) f^{pq}\notag\\
-\frac{1}{20}{Sym}\Bigl(
        &( 8 a_{iqp}  f_{rk} - 42 a_{pqk}  f_{ri} ) f_{sj} a_{mlt} + (3 a_{tm{\mu}} a_{rs{\nu}} - 4 a_{{\mu}mt} a_{{\nu}rs} ) f_{ip} f_{jq} f_{kl}f^{{\mu}{\nu}}+\notag\\
        &(64 a_{kqp} f_{sj} - 12 a_{kpq} f_{sj} + 10 a_{pqj}  f_{sk} - 20 a_{jqp}  f_{sk} ) a_{ltm} f_{ri} +\notag\\
        &(102 a_{iqp} a_{tmj} - 51 a_{pqi} a_{tmj} - 48 a_{iqp} a_{jmt} ) f_{rk} f_{sl}+\\
        &(28 a_{kqp} a_{imt}  - 32 a_{kpq} a_{imt} + 10 a_{pqk} a_{tmi} ) f_{rj} f_{sl}+\notag\\
        &(42 a_{pqi} f_{rk} f_{sj} - 40 a_{ipq}  f_{rk} f_{sj} + 20 a_{jqp}  f_{ri} f_{sk} ) a_{tml} +\notag\\
        &(32 a_{{\mu}tm} a_{{\nu}rs} -34 a_{{\mu}tm} a_{rs{\nu}} ) f_{kp} f_{jq} f_{il}f^{{\mu}{\nu}}\notag
        \Bigr) \varepsilon^{ptr} \varepsilon^{qms},
\end{align}
here $Sym$ denotes  symmetrization with respect to $i$ and $j$. Both conditions (\ref{int1}), ({\ref{int2}) simplify to (\ref{Lint1}), ({\ref{Lint2}) under the Lagrangian assumption.
These conditions provide a straightforward computer test of the integrability for equations from the class
(\ref{u}).

\subsection{Integrable equations and conformal structures possessing conjugate null coordinate systems}

Our first result  is the following

\begin{theorem}
The conformal structure $f_{ij}dp^idp^j $ corresponding to any integrable three-dimensional equation (\ref{u}) is conformally flat.
\end{theorem}

The proof is a straightforward  calculation of the corresponding Cotton tensor, based on the integrability conditions derived in Sect. 2. Recall that for three-dimensional Lagrangian systems this result was established earlier in \cite{FKT}. We emphasize that the transformation  which brings the metric to  a constant coefficient form is not necessarily projective (it does become projective for linearizable systems only). Thus, the theory of integrable equations of the form (\ref{u}) has two `flat'  counterparts:
the first one is a flat projective structure provided by the projective space $P^n$ with coordinates $p^1, ..., p^n$,  and the standard projective action of $SL(n+1, R)$. The second is the flat conformal structure  $f_{ij}dp^idp^j$. Although, viewed separately, both structures are trivial, this is no longer true when they are imposed simultaneously: their  `flat coordinate systems'  do not coincide in general.

\medskip

Our next goal is to provide a differential-geometric characterization of hydrodynamic reductions. Although our discussion will be restricted to the dimension three, all  conclusions hold in any dimension. We will follow the notation of Sect. 2. Let ${\bf p}=(p^1, p^2, p^3)=(a, b, c)$  be functions of the Riemann invariants $R^i$ (for our purposes it will be sufficient to consider one-, two- and three-component reductions only). By virtue of (\ref{bc}), the derivative of ${\bf p}$ with respect to $R^i$ is given by
$$
\partial_i{\bf p}=(1, \mu^i, \lambda^i)\ \partial_ia.
$$
The dispersion relation (\ref{disp}) implies that $\partial_i{\bf p}$ a null vector of the conformal structure $f_{ij}dp^idp^j$. Thus, the Riemann invariants $R^i$ provide a net of null curves on the corresponding submanifold ${\bf p}({\bf R})$. Furthermore, the relation
$$
\partial_i\partial_j{\bf p}\in {\rm span} \{\partial_i{\bf p}, \partial_j{\bf p}\},
$$
which readily follow from (\ref{comm}), implies that this net is conjugate. Thus, we have the following geometric picture:

\noindent {\bf One-component reductions} correspond to null curves of the associated conformal structure.

\noindent {\bf Two-component reductions} are in one-to-one correspondence with surfaces  which carry a conjugate net of null curves. Notice that we are in the realm of two different geometries, namely, conformal geometry (responsible for the property of being null), and projective geometry (responsible for the property of  being conjugate).

\noindent {\bf Three-component reductions} correspond to three-conjugate null coordinate systems  in $P^3$. Since the existence of `sufficiently many' three-component reductions is a necessary and sufficient condition for the integrability,  the problem of the classification of three-dimensional integrable equations of the form (\ref{u}) can be reformulated geometrically as follows:
{\it classify conformal structures in the projective space $P^3$ which possess infinitely many three-conjugate null coordinate systems  parametrized by three arbitrary functions of one variable}.

It is a truly remarkable fact that the moduli space of such structures is only $20$-dimensional!

\section{Appendix: proof of Theorem 1}

Here we provide further details of the proof which was only sketched in Sect. 2.
Our starting point is  the quasilinear representation of Eq. (\ref{u}),
\begin{equation}
a_y=b_x, ~~~ a_t=c_x, ~~~ b_t=c_y, ~~~
f_{11} a_x+ f_{22} b_y + f_{33} c_t+ 2f_{12} a_y  + 2f_{13} a_t  + 2f_{23} b_t=0.
\label{1}
\end{equation}
Following the method of hydrodynamic reductions, we seek multi-phase solutions  in the form
\begin{equation}
a=a(R^1, \ldots ,
R^N), \  \ b=b(R^1, \ldots , R^N), \ \ c=c(R^1, \ldots , R^N),
\label{an}
\end{equation}
 where the  phases
$R^1(x,y,t)$, \ldots , $R^N(x,y,t)$ are {\it arbitrary} solutions of
Eqs.  (\ref{R}).
Substituting the ansatz (\ref{an}) into (\ref{1}) one obtains the equations
\begin{equation}
\partial_i b=\mu^i\partial_i a, ~~~ \partial_i c =\lambda^i\partial_i a,
\label{bc}
\end{equation}
along with the dispersion relation
\begin{equation}
D(\lambda^i, \mu^i) =
f_{11} + f_{22}(\mu^i)^2 + f_{33} (\lambda^i)^2 + 2f_{12}\mu^i + 2f_{13} \lambda^i
 + 2f_{23}\mu^i\lambda^i=0.
\label{disp}
\end{equation}
Hereafter, we assume the conic (\ref{disp}) to be irreducible. This condition is
equivalent to the non-vanishing of the determinant of the coefficient matrix
\begin{equation*}
F = \left(
\begin{array}{ccc}
f_{11} & f_{12} & f_{13}\\
f_{12} & f_{22} & f_{23}\\
f_{13} & f_{23} & f_{33}
\end{array}
\right), ~~~ {\rm det} \  F\ne 0.
\end{equation*}
The consistency conditions for Eqs. (\ref{bc}) imply
\begin{equation}
\partial_i\partial_ja=\frac{\partial_j\lambda
^i}{\lambda^j-\lambda^i}\partial_i
a+\frac{\partial_i\lambda
^j}{\lambda^i-\lambda^j}\partial_j a.
\label{a}
\end{equation}
Differentiating the dispersion relation (\ref{disp}) with respect to $R^j, \
j\ne i,$ and keeping in mind Eqs.
(\ref{bc}) and (\ref{comm}), one obtains  explicit expressions for $\partial_j
\lambda^i$ and
$\partial_j \mu^i$ in the form
\begin{equation}
\partial_j\lambda^i=(\lambda^i-\lambda^j)B_{ij}\partial_ja, ~~~
\partial_j\mu^i=(\mu^i-\mu^j)B_{ij}\partial_ja,
\label{2}
\end{equation}
where $B_{ij}$ are  {\it rational} expressions in $\lambda^i, \lambda^j, \mu^i, \mu^j$
whose coefficients depend on $f_{ij}(a,b,c)$ and  first order
derivatives thereof. Explicitly, one has
$$
B_{ij}=\frac{N_{ij}}{D_{ij}}=\frac{1}{2} \frac{N_{ij}}
{f_{11}+f_{22}\mu^i\mu^j+f_{33}\lambda^i\lambda^j+f_{12}(\mu^i+\mu^j)+f_{13}(\lambda^i+\lambda^j)+
f_{23}(\mu^i\lambda^j+\mu^j\lambda^i)};
$$
notice that,  modulo the dispersion relation (\ref{disp}),  the denominator
$D_{ij}$ equals
$4D\left(\frac{\lambda^i+\lambda^j}{2}, \ \frac{\mu^i+\mu^j}{2}\right)$.
The numerator $N_{ij}$ is a polynomial expression of the form
$$
\begin{array}{c}
N_{ij} =
      f_{33,3}{\lambda^{i}}^2{\lambda^{j}} + f_{33,2}{\mu^{j}}{\lambda^{i}}^2 +
      f_{33,1}{\lambda^{i}}^2 + 2f_{23,3}{\mu^{i}}{\lambda^{i}}{\lambda^{j}} + 2
      f_{23,2}{\mu^{i}}{\mu^{j}}{\lambda^{i}} + 2f_{23,1}{\mu^{i}}{\lambda^{i}}\\
\ \\
       + f_{22,3}{\mu^{i}}^2{\lambda^{j}} + f_{22,2}{\mu^{i}}^2{\mu^{j}} +
      f_{22,1}{\mu^{i}}^2 + 2f_{13,3}{\lambda^{i}}{\lambda^{j}} + 2f_{13,2}
      {\mu^{j}}{\lambda^{i}} + 2f_{13,1}{\lambda^{i}}\\
\ \\
      + 2f_{12,3}{\mu^{i}}{\lambda^{j}} + 2f_{12,2}{\mu^{i}}{\mu^{j}} + 2f_{12,1}{\mu^{i}} +
      f_{11,3}{\lambda^{j}} + f_{11,2}{\mu^{j}} + f_{11,1};
\end{array}
$$
we adopt the convention that variables $a, b, c$ correspond to indices $1, 2, 3$:  thus, $f_{11,1}=f_{11, a}$, etc. Taking into account Eqs. (\ref{2}), Eqs. (\ref{a}) assume the form
\begin{equation}
\partial_i\partial_ja=-(B_{ij}+B_{ji})\partial_ia\partial_ja.
\label{a1}
\end{equation}
The compatibility conditions $\partial_k\partial_j
\lambda^i=\partial_j\partial_k \lambda^i$,
 $\partial_k\partial_j \mu^i=\partial_j\partial_k \mu^i$ and
$\partial_k\partial_j\partial_i a=\partial_j\partial_k\partial_i a$ are
equivalent to the equations
\begin{equation}
\partial_kB_{ij}=(B_{ij}B_{kj}+B_{ij}B_{ik}-B_{kj}B_{ik})\partial_ka,
\label{int}
\end{equation}
which must be satisfied identically by virtue of Eqs. (\ref{bc}), (\ref{disp}),
(\ref{2}) and (\ref{a1}).  In order to obtain equations with `simplest possible'  coefficients at the second order derivatives of $f_{ij}(a,b,c)$ we rewrite Eqs. (\ref{int}) as
\begin{equation}
\label{Nij}
\partial_kN_{ij}=N_{ij}\left(\frac{1}{D_{ij}}\partial_kD_{ij} +
B_{kj}\partial_ka+B_{ik}\partial_ka\right)-D_{ij}B_{kj}B_{ik}\partial_ka.
\end{equation}
The second order derivatives of $f_{ij}(a,b,c)$ are present only in
the l.h.s.\ term $\partial_kN_{ij}$. Further reduction of the complexity
of the expression in the r.h.s.\ is achieved by representing
$1/D_{ij}$ in the form
$$
\begin{array}{c}
\displaystyle \frac{1}{D_{ij}} = U_{ij}=
[2(\lambda^if_{23}+f_{12})(\lambda^jf_{23}+f_{12})
  -f_{22}(\lambda^i\lambda^jf_{33}
 + (\lambda^i+\lambda^j)f_{13} +f_{11}) \\
  \ \\
+f_{22}(\lambda^jf_{23}+f_{12})\mu^i
+f_{22}(\lambda^if_{23}+f_{12})\mu^j
+f_{22}^2\mu^i\mu^j ]
/( 2 (\lambda^i- \lambda^j)^2{\rm det}F),
\end{array}
$$
which holds identically modulo the dispersion relation (\ref{disp}),
and a subsequent substitution $B_{st}=N_{st}/D_{st} = N_{st}U_{st}$.
The denominators of the r.h.s.\ terms in Eqs. (\ref{Nij}) cancel out,
producing
a polynomial in  $\lambda^i$, $\lambda^j$, $\lambda^k$, $\mu^i$,
$\mu^j$, $\mu^k$ with coefficients depending on the functions $f_{ij}(a,b,c)$
and their derivatives up to second order.
This was the most essential technical part of the calculation:
the original expression  (\ref{Nij}) has more than 1.000.000
terms with different denominators;
after properly organized cancellations it reduces to a polynomial expression with less
than 4500 terms.
Using the dispersion relation (\ref{disp}) and assuming, say,
$f_{22} \neq 0$ (this can always be achieved by a linear change of the independent
variables $x, y, t$), we simplify this polynomial by excluding
the powers  $(\mu^i)^s$,
$(\mu^j)^s$, $(\mu^k)^s$,  $s\geq 2$,
arriving at a polynomial of degree one in each of $\mu^i$,
$\mu^j$, $\mu^k$, and degree two in $\lambda$'s.
Equating similar coefficients of these polynomials in
both sides of Eqs. (\ref{Nij}),
we arrive at a set of 45 equations for the derivatives of the coefficients
$f_{ij}(a, b, c)$, which are linear in the second order derivatives. One can verify that only   30 of these equations are  linearly
independent.
Solving them, we get closed form expressions for the second order partial derivatives
of the coefficients  $f_{11}, f_{12}, f_{13}, f_{23}, f_{33}$ in terms of the first
order derivatives thereof, $30$ equations altogether (without any loss of generality one can set  $f_{22} = 1$), which can be represented in symbolic form (\ref{int-cond}),
$$
d^2f_{ij}=\frac{1}{F}R(f_{kl}, df_{kl});
$$
here $R$ is quadratic in both $f_{kl}$ and $df_{kl}$. A straigtforward computation shows that this system is in involution: all compatibility conditions
are satisfied identically. Since the values of the five functions $f_{11}, f_{12}, f_{13}, f_{23}, f_{33}$, and first order derivatives thereof, are not restricted by any additional constraints, we obtain a $5+3\cdot 5=20$-dimensional moduli space of integrable equations. This finishes the proof of Theorem 1.




\section*{Acknowledgements}

We thank B Dubrovin, K Khusnutdinova, M Pavlov and A Veselov for their interest, constant support  and valuable remarks. EVF thanks A Odesskii for clarifying discussions on modular forms. The research of EVF and PAB  was partially supported by the EPSRC grant EP/D036178/1,  the
European Union through the FP6 Marie Curie RTN project ENIGMA (Contract
number MRTN-CT-2004-5652), and the ESF programme MISGAM. SPT is grateful to the Institute of Mathematics in Taipei (Taiwan) where a
part of this work has been completed, and especially to Jen-Hsu Chang for the
hospitality at the National Defense University. The research of SPT was partially
supported by the Russian--Taiwanese grant
06-01-89507-HHC (95WFE0300007), and the RFBR grant 06-01-00814.


\begin{thebibliography}{99}

\bibitem{A2} M.J. Ablowitz, S. Chakravarty and H. Hahn, Integrable systems and modular forms of level $2$, J. Phys. A: Math. Gen. {\bf 39} (2006) 15341--15353.

\bibitem {Adler} V.E. Adler and  A.B. Shabat,
Model equation of the theory of solitons, {\bf 153}, no. 1 (2007) 1373--1387.


\bibitem{Akivis} M.A. Akivis and V.V. Goldberg, Projective differential geometry of submanifolds, Elsevier Science Publishers (1993) 362pp.


\bibitem{Avritzer} D. Avritzer and H. Lange, Moduli spaces of quadratic complexes and their singular surfaces, Geom. Dedicata {\bf 127} (2007) 177--197.



\bibitem{Bla} M. Blaszak, B.M. Szablikowski, Classical R-matrix
theory of
dispersionless systems: II. (2+1)-dimension theory, J. Phys. A {\bf 35} (2002)
10345--10364.

\bibitem{Bla1} M. Blaszak, Classical R-matrices on Poisson algebras and related dispersionless systems, Phys. lett A {\bf 297} (2002) 191--195.

\bibitem{Bogdanov} L.V. Bogdanov, B.G. Konopelchenko and L.
Martinez Alonso, Quasi-classical $\overline\partial$-method: Generating
equations for dispersionless integrable hierarchies, Teoret. Mat. Fiz. {\bf 134}
(2003)  46--54.


\bibitem{Bogdanov1} L.V. Bogdanov and B.G. Konopelchenko, Nonlinear Beltrami equations and $\tau$-functions for dispersionless  hierarchies, Phys. Letters A:  {\bf 322}, no. 5-6 (2004) 330--337.

\bibitem{Bogdanov2} L.V. Bogdanov and B.G. Konopelchenko, On dispersionless BKP hierarchy and its reductions, J. Nonlinear Math. Phys. {\bf 12}, suppl. 1 (2005)  64--73.

\bibitem{BF} {C.P. Boyer} and{\ J.D. Finley,} Killing vectors in
self-dual Euclidean Einstein spaces, J. Math. Phys. \textbf{23} (1982)
1126--1130.

\bibitem{Kodama} R. Carroll and Y. Kodama,  Solution of the dispersionless Hirota equations, J. Phys. A {\bf 28}, no. 22 (1995) 6373--6387.


\bibitem{Chazy} {J. Chazy,} Sur les \'{e}quations diff\'{e}rentiellles
dont l'int\'{e}grale g\'{e}n\'{e}rale poss\`{e}de un coupure essentielle
mobile, C.R. Acad. Sc. Paris, \textbf{150} (1910) 456--458.


\bibitem{Dub} B.A. Dubrovin and S.P. Novikov,
Hydrodynamics of weakly
deformed soliton lattices:
differential geometry and
Hamiltonian theory, Russian Math. Surveys,
{\bf 44}  (1989) 35--124.





\bibitem{D} M. Dunajski, L.J. Mason and P. Tod,
Einstein-Weyl geometry, the dKP equation and twistor theory,
J. Geom. Phys. {\bf 37}, no. 1-2 (2001) 63--93.



\bibitem{F} E.V. Ferapontov, D.A. Korotkin  and V.A. Shramchenko, Boyer-Finley equation and systems of hydrodynamic type, Class. Quantum Grav. {\bf 19}, no. 24 (2002) L205--L210.

\bibitem{Fer22} E.V. Ferapontov and  M.V. Pavlov, Hydrodynamic reductions of the heavenly equation,
Class. Quantum Grav. {\bf 20} (2003) 2429--2441.





\bibitem{Fer4} E.V. Ferapontov and K.R. Khusnutdinova, On integrability of
(2+1)-dimensional quasilinear systems, Comm. Math. Phys.  {\bf 248} (2004)
187--206.

\bibitem{Fer5} E.V. Ferapontov and K.R. Khusnutdinova, The characterization of
two-component (2+1)-dimensional integrable  systems
of hydrodynamic type,   J. Phys. A: Math. Gen. {\bf 37}  (2004)
2949--2963.


\bibitem{FKP} E.V. Ferapontov, K.R. Khusnutdinova and
M.V. Pavlov, On the classification of integrable equations of
the form $\Omega_{tt}=f(\Omega_{xx}, \Omega_{xt}, \Omega_{xy})$,
Theor. Math. Phys. {\bf 144} (2005)  35--43.

\bibitem{FKT} E.V. Ferapontov, K.R. Khusnutdinova and
S.P. Tsarev, On a class of three-dimensional integrable Lagrangians, Comm. Math. Phys. {\bf 261}, N1 (2006)  225--243.


\bibitem{Lenos} E.V. Ferapontov, L. Hadjikos and K.R. Khusnutdinova, Integrable equations of the dispersionless Hirota type and hypersurfaces in the Lagrangian Grassmannian, arXiv: 0705.1774 (2007).

\bibitem{Odesskii} E.V. Ferapontov and  A.V. Odesskii,
Integrable Lagrangians and modular forms, arXiv:0707.3433, (2007).


\bibitem{Moro} E.V. Ferapontov, A. Moro and  V.V. Sokolov,
Hamiltonian systems of hydrodynamic type in 2+1 dimensions, arXiv:0710.2012v1, (2007).


\bibitem{Gibb94} J. Gibbons and Y. Kodama,   A method for solving the
dispersionless KP hierarchy and its exact solutions. II. Phys. Lett.
A {\bf135} (1989) 167--170.

\bibitem{GibTsa96} J. Gibbons and S.P. Tsarev,
Reductions of the
Benney equations, Phys. Lett. A {\bf 211} (1996) 19--24.

\bibitem{GibTsa99} J. Gibbons and S.P. Tsarev, Conformal maps and
reductions of
the Benney equations, Phys. Lett. A {\bf 258} (1999)
263--271.




\bibitem{Hitchin} N. J. Hitchin, Complex manifolds and Einstein's equations,  in  Twistor geometry and nonlinear systems, Lecture Notes in Math. {\bf 970} Springer, Berlin-New York (1982) 73--99.

\bibitem{Jessop} C.M. Jessop,  A treatise on the line complex, Chelsea Publishing Co., New York (1969)  364 pp.

\bibitem{Ko} B. Konopelchenko, Soliton eigenfunction equations: the IST integrabilty
and some properties, Reviews in Math.Physics {\bf  2} (1990) 399--440.

\bibitem{Kon} B.G. Konopelchenko and L. Martinez Alonso,
Dispersionless scalar integrable hierarchies, Whitham
hierarchy, and the quasiclassical $\overline\partial$-dressing method, J. Math.
Phys. {\bf 43}, no. 7 (2002) 3807--3823.

\bibitem{Kr1} I.M. Krichever, The averaging method for
two-dimensional "integrable" equations, Funct. Anal. Appl. {\bf 22}, no. 3
(1988) 200--213.



\bibitem{Kr} I.M. Krichever,  The $\tau$-function of the universal Whitham hierarchy, matrix models and topological field theories, Comm. Pure Appl. Math. {\bf 47}, no. 4 (1994) 437--475.


\bibitem{Kr3} I.M. Krichever, M. Mineev-Weinstein, P. Wiegmann and A. Zabrodin,
Laplacian growth and Whitham equations of soliton theory, Phys. D {\bf 198}, no. 1-2 (2004) 1--28.


\bibitem{Kup1} B.A. Kupershmidt, Geometric Hamiltonian forms for the Kadomtsev-Petviashvili and Zabolotskaya-Khokhlov equation, in Geometry in Partial Differential Equations, eds. A. Prastaro and Th. M. Rassias, World Scientific Publishing Co. (1994) 155--172.

\bibitem{Manakov} S.V. Manakov and P.M.  Santini, On the solutions of the dKP equation: the nonlinear Riemann Hilbert problem, longtime behavior, implicit solutions and wave breaking, arXiv:0707.1802, to appear in J. Phys. A: Math. Theor {\bf 41} (2008).


\bibitem{Manas} M. Manas, On the $r$th dispersionless Toda hierarchy: factorization problem, additional symmetries and some solutions, J. Phys. A {\bf 37}, no. 39 (2004)  9195--9224.

\bibitem{Ma} M. Manas,  L. Martinez
Alonso and E. Medina,
Reductions and hodograph solutions of the dispersionless
KP
hierarchy, J. Phys. A: Math. Gen. {\bf 35} (2002) 401--417.





\bibitem{Shabat} L. Martinez Alonso and A.B. Shabat, Hydrodynamic reductions and
solutions of a universal hierarchy, Teoret. Mat. Fiz. {\bf 140} (2004), 216--229.

\bibitem{O} A.V. Odesskii,  A family of (2+1)-dimensional hydrodynamic-type systems possessing
pseudopotential, arXiv:0704.3577v3.

\bibitem{Ovsienko} V. Ovsienko,
Bi-Hamiltonian nature of the equation $u_{tx}=u_{xy} u_y-u_{yy} u_x$, arXiv:0802.1818.





\bibitem{Pavlov1} M.V. Pavlov, Classifying integrable
Egoroff hydrodynamic chains,  Theoretical and Mathematical Physics,
{\bf 138} (1) (2004) 45--58.

\bibitem{Safaryan}
L.P. Safaryan,  Certain classes of manifolds of cones of order two in $P\sb{n}$. (Russian) Akad. Nauk Armjan. SSR Dokl. {\bf 50} (1970) 83--90.


\bibitem{Sidorov} A.F. Sidorov, V.P. Shapeev and N.N. Yanenko, The method of
differential constraints and its applications in gas dynamics, ``Nauka'',
Novosibirsk (1984) 272 pp.


\bibitem{Tsarev} S.P. Tsarev, Geometry of Hamiltonian systems of
hydrodynamic
type. Generalized hodograph method, Izvestija AN USSR
Math. {\bf 54}  (1990)
1048--1068.

\bibitem{KZ} E.A. Zabolotskaya and R.V. Khokhlov, Quasi-plane waves in the nonlinear acoustics of confined beams, Sov. Phys. Acoust. {\bf 15} (1969) 35--40.

\bibitem{Zakharevich} I. Zakharevich
Nonlinear wave equation, nonlinear Riemann problem, and the twistor transform of Veronese webs, arXiv:math-ph/0006001.


\bibitem{Zakharov}  V.E. Zakharov,  Dispersionless limit of
integrable systems
in $2+1$ dimensions, in Singular Limits of
Dispersive Waves, Ed. N.M. Ercolani
et al., Plenum Press, NY (1994)
165--174.


\bibitem{FORM}
  J.A.M. Vermaseren "New features of FORM" arXiv:math-ph/0010025,
a complete distribution can be  downloaded from
{\tt http://www.nikhef.nl/{\~{}}form/}

\bibitem{Maple}
Maple 9.5 Getting Started Guide.
Toronto: Maplesoft, a division of Waterloo Maple Inc., 2004.

\end{thebibliography}
\end{document}